\documentclass[onefignum,onetabnum]{siamart190516}

\usepackage[latin1]{inputenc}
\usepackage[T1]{fontenc}
\usepackage[english]{babel}
\usepackage{amssymb}
\usepackage{amsmath}
\usepackage{amsfonts}
\usepackage{amstext}
\usepackage{soul}
\usepackage{psfrag}
\usepackage{epsfig}
\usepackage{bbm}
\usepackage{bm}
\usepackage{bbold}
\usepackage{calc}
\usepackage{mathrsfs}
\usepackage{textcomp}
\usepackage{color,hyperref}

\usepackage{array}
\usepackage{booktabs}
\usepackage{multirow} 

\usepackage{lipsum}
\usepackage{graphicx}
\usepackage{algorithmic}
\ifpdf
  \DeclareGraphicsExtensions{.eps,.pdf,.png,.jpg}
\else
  \DeclareGraphicsExtensions{.eps}
\fi

\newsiamremark{remark}{Remark}
\newsiamremark{hypothesis}{Hypothesis}
\crefname{hypothesis}{Hypothesis}{Hypotheses}
\newsiamthm{claim}{Claim}

\headers{Motifs for processes on networks}{A. C. Schwarze, M. A. Porter}

\title{Motifs for processes on networks\thanks{Submitted to the editors May 10, 2021.
\funding{A.C.S. was supported by the Engineering and Physical Sciences Research Council under grant number EP/L016044/1, the Clarendon Fund, and e-Therapeutics plc. M.A.P. acknowledges support from the National Science Foundation (grant number 1922952) through the Algorithms for Threat Detection (ATD) program.}}}

\author{Alice C. Schwarze\thanks{Department of Biology, University of Washington, Seattle, WA
  (\email{schwarze@uw.edu}).}
\and Mason A. Porter\thanks{Department of Mathematics, University of California Los Angeles, Los Angeles, CA 
  (\email{mason@math.ucla.edu}).}
}

\usepackage{amsopn}

\ifpdf
\hypersetup{
  pdftitle={Motifs for processes on networks},
  pdfauthor={A. C. Schwarze, M. A. Porter}
}
\fi

\newcommand\covmat{{\bf\Sigma}}
\newcommand\corrmat{{\bf R}}
\newcommand\covel{\sigma}
\newcommand\correl{r}
\newcommand\specific{\hat c}
\newcommand\effi{\eta}
\newcommand\cvec{\phi}

\newcommand\kOne{{k_1}}
\newcommand\kTwo{{k_2}}
\newcommand\lastIndex{k'}

\newcommand{\norm}[1]{\left\lVert#1\right\rVert}
\newcommand\indA{{m'}} 
\newcommand\indB{{k}} 

\newcommand{\panel}[1]{(\lowercase{#1})} 

\newcommand{\eq}[1]{\cref{eq:#1}}

\begin{document}

\maketitle

\begin{abstract}
The study of motifs in networks can help researchers uncover links between the structure and function of networks in biology, sociology, economics, and many other areas. Empirical studies of networks have identified feedback loops, feedforward loops, and several other small structures as ``motifs'' that occur frequently in real-world networks and may contribute by various mechanisms to important functions in these systems. However, these mechanisms are unknown for many of these motifs. We propose to distinguish between ``structure motifs'' (i.e., graphlets) in networks and ``process motifs'' (which we define as structured sets of walks) on networks and consider process motifs as building blocks of processes on networks. Using the steady-state covariances and steady-state correlations in a multivariate Ornstein--Uhlenbeck process on a network as examples, we demonstrate that the distinction between structure motifs and process motifs makes it possible to gain quantitative insights into mechanisms that contribute to important functions of dynamical systems on networks.
\end{abstract}





\section {Introduction}

The study of motifs in networks has advanced the understanding of various systems in biology \cite{Alon2007, Rip2010, Ristl2014, Shen-Orr2002, Stone2019}, economics \cite{Ohnishi2010, Takes2018}, social science \cite{Hong-Lin2014, Juszczyszyn2012}, and other areas. When interpreting motifs as small building blocks that can contribute to a network's functionality, it can be important to identify motifs that are necessary, beneficial, or disadvantageous to a network's function to help uncover the relationship between network structure and network function.

Traditionally, scientists have considered graphlets (i.e., small graphs of typically three to five nodes) as building blocks of a network's structure and identified them as ``motifs'' when empirical data \cite{Conant2003, Hong-Lin2014, Milo2002, Ohnishi2010, Shen-Orr2002,  Sporns2004, Takes2018} or mathematical models \cite{Antoneli2018, Golubitsky2009, Ingram2006, Shilnikov2008} indicate their importance to system function. In many studies of ``real-world'' networks from empirical data, researchers have compared graphlet frequencies in a network to graphlet frequencies in an appropriate random-graph null model \cite{Conant2003, Hong-Lin2014, Milo2002, Ohnishi2010, Sporns2004, Takes2018}. They subsequently have concluded that graphlets that are overrepresented in the network are likely to be relevant for important functions of the system that is associated with that network. However, the results of such studies depend very sensitively on the choice of an appropriate random-graph null model \cite{Artzy-Randrup2004, Robin2007, Schlauch2015}, and this approach to motif identification does not uncover the mechanisms by which the identified graphlets contribute to important system functions. 

Other studies have aimed to provide mechanistic insights by modeling dynamical systems on graphlets in isolation \cite{Antoneli2018, Golubitsky2009, Ingram2006, Shilnikov2008}. The design of such studies requires an \textit{a priori} choice of a graphlet, a dynamical system or a class of dynamical systems, and a candidate mechanism by which the graphlet facilitates an important system function. Therefore, it is difficult for such studies to discover new and/or unexpected mechanisms or to provide a systematic comparison of the importances of different graphlets and different mechanisms for a system function.

In the present paper, we propose a framework for connecting the study of dynamics on networks with the study of motifs in networks.
We propose to distinguish between ``structure motifs'' (i.e., graphlets) in networks and ``process motifs'' (which we define in the form of structured sets of walks) on networks, and we consider process motifs as building blocks for processes on networks\footnote{In other studies, the term ``structural motif'' often has been used to refer to structure motifs, but it sometimes has been used to refer to process motifs. We give an overview of the use of motifs in the study of networks in \cref{sec:litreview}. We use the terms ``structure motif'' and ``process motif'' to avoid confusion with conflicting definitions of ``structural motif'' in previous work by other scholars. We use the composite nouns to stress that we consider structure motifs and process motifs to be two fundamentally different concepts.}. We demonstrate how to use process motifs to connect network structure to dynamics on networks and to dynamics-based notions of system functions. These connections lead to mechanistic and quantitative insights into the contribution of all possible structure motifs to a given system function. We give concrete examples in \cref{sec:example}.

\begin{figure*}[t] 
\centering
  \includegraphics[trim={3.2cm 1.3cm 2.9cm 1cm},clip, width=1\textwidth]{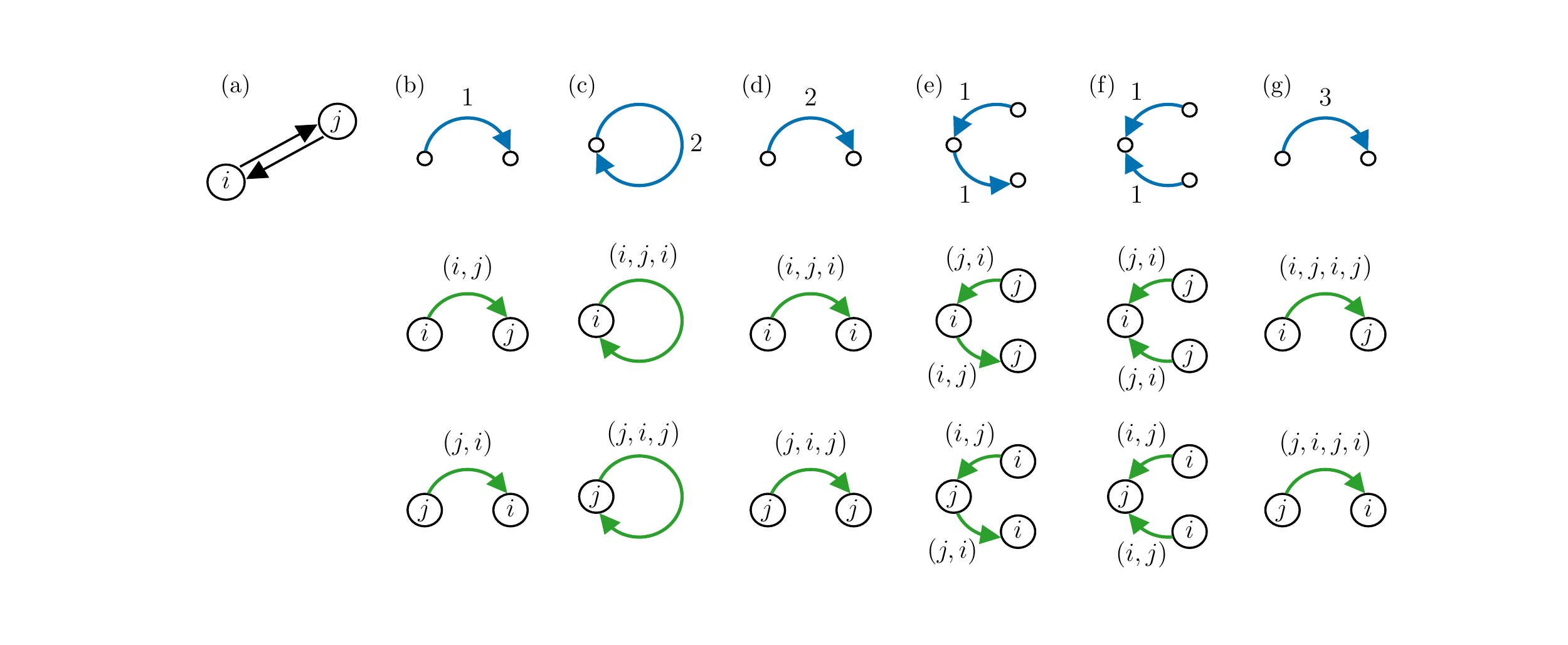}  
\caption{Examples of process motifs. In \panel{A}, we show a small directed network $(V,E)$. In \panel{B}--\panel{G}, small white nodes and blue curved edges depict examples of process motifs that occur on $(V,E)$. Numerical edge labels indicate the length of a walk. There are two occurrences of each process motif
on $(V,E)$. We use labeled white nodes and curved green edges to depict these occurrences below each process motif.
}
\label{fig:walk_graphs}
\end{figure*}

\begin{figure*}[t] 
\centering
  \includegraphics[trim={4.3cm 15.75cm 3.9cm 0.75cm},clip, width=1\textwidth]{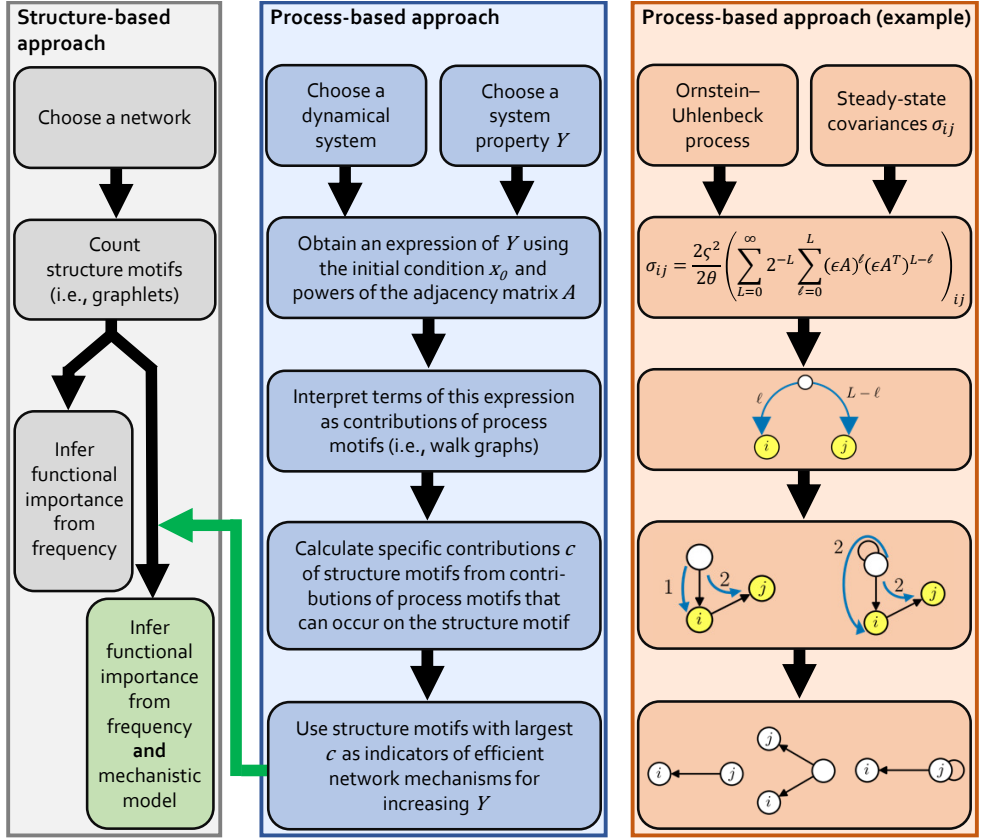}  
  \caption{Comparison of a structure-based approach to the study of motifs in networks and the process-based approach that we introduce in this paper. In the right panel, we give an overview of the results of applying our approach to the study of process motifs and structure motifs that are relevant for steady-state covariances $\covel_{ij}$ (see \eq{sigma}) for node pairs $(i,j)$ in a multivariate Ornstein--Uhlenbeck process with parameters $\theta$, $\varsigma$, and $\epsilon$ and adjacency matrix ${\bf A}$ (see \eq{OU}). The parameters $L$ and $\ell$ characterize a process motif for $\covel_{ij}$ with walk lengths $\ell$ and $L-\ell$.
  }
  \label{fig:flow}
\end{figure*}

We define a \textit{process motif} to be a connected \textit{walk graph}, which we define to be a directed and weighted multigraph in which each edge corresponds to a walk on a network. In line with prior research on motifs, we are concerned with small process motifs.
Edge weights in a walk graph correspond to the lengths of the associated walks. 
An \textit{occurrence} of a process motif on a network $(V,E)$ with node set $V$ and edge set $E$ is a labeling of nodes and edges in the process motif such that each node in the process motif corresponds to a node in $V$ and each edge in the process motif corresponds to a walk on $(V,E)$. (This technical notion of ``occurrence'' is consistent with the English meaning of the word.)
In \cref{fig:walk_graphs}, we show examples of process motifs and their occurrences on a small network. The occurrences of the process motifs in panels \panel{B} and \panel{C} use each edge in $E$ at most once, and each node in these occurrences corresponds to a different node in $V$. In the occurrences of the process motifs in panels \panel{D}--\panel{F}, some nodes correspond to the same node in $V$. In the occurrences of the process motifs in panels \panel{F} and \panel{G}, walks use edges in $E$ more than once. 

In \cref{fig:flow}, we give an overview and an example of our process-based approach to studying motifs in networks; we also indicate how our results can inform future studies of motifs in network structure. We model a system function as a real-valued function $Y$ of the state of a dynamical system. One can identify the process motifs that are relevant for a given mathematical function $Y$ and associate each process motif with a numerical value $b$ that indicates the contribution of each of the process motif's occurrences to $Y$. From process motifs and their associated contributions, one can derive structure motifs that are relevant to the function $Y$ and the contributions $c$ of their occurrences to $Y$. Process motifs thus offer a framework for identifying functionally important graphlets (i.e., structure motifs) from mathematical models. This approach can lead to detailed insights into the mechanisms by which structure motifs can affect a system function (see \cref{sec:example})~\footnote{Note that we distinguish a ``system function'' (which may, for example, be a biological function in a system) from a ``mathematical function'' like $Y$.}.  One can use contributions that are associated with structure motifs to rank mechanisms based on their efficiency and thereby rank structure motifs based on their importance in contributing to a system function. 

As an example system, we use the multivariate Ornstein--Uhlenbeck process \linebreak (mOUP), which is a popular model for noisy coupled systems \cite{Aalen2004}. It has been applied to study neuronal dynamics \cite{Barnett2009}, stock prices \cite{Liang2011}, gene expression \cite{Rohlfs2013}, and other systems. Properties of the mOUP are related to properties of coupled excitable systems. For example, one can derive the mOUP as a linear-response approximation of an integrate-and-fire model for excitable neurons \cite{Grytskyy2013, Hu2014}.

As example system properties, we examine the covariances and the correlations in the mOUP at steady state. Covariances and correlations between pairs of nodes in a network are relevant for a wide variety of topics.
Researchers have used correlations between variables to construct networks for various applications \cite{Carter2004, Fox2010, Onnela2004}. For example, in networks of functional connectivity, an edge may indicate a large positive correlation between two neurons or two brain regions \cite{Fox2010}. In networks of gene co-expression, an edge may indicate a strong correlation between the expression of two genes \cite{Carter2004}. Additionally, existing intuitive results on simple network structures that induce covariance and correlation (see, e.g., Reichenbach's common-cause principle \cite{Reichenbach1956}) make covariance and correlation interesting examples for our study. 
Our approach confirms known results about covariation between variables and yields additional, quantitative insights into the mechanisms by which network structure can enhance or reduce covariance or correlation between nodes.

Our process-based approach to the study of motifs on networks yields a list of relevant process motifs (with their associated contributions to a system function) and a list of relevant structure motifs (with their associated contributions to the same system function). As we indicate in \cref{fig:flow}, these results depend both on the choice of dynamical system and on the choice of system function. However, they do not depend on the choice of network or random-graph model. In \cref{fig:flow}, the arrow from the center panel to the left panel indicates how our results can inform future studies of graphlets in networks and can lead to quantitative insights into the importances of graphlets for a system function in a random-graph model or in a given network (from data or from a random-graph model).

Our paper proceeds as follows. In \cref{sec:definitions}, we review some graph-theoretical concepts and define walk graphs. The concept of walk graphs allows us
to distinguish between structure motifs and process motifs. We also provide an overview of the use of motifs in prior studies of networks. In \cref{sec:part2}, we show how to derive process motifs, structure motifs, and their contributions to a given property (such as a correlation) of a dynamical system. In \cref{sec:example}, we give a brief introduction to the mOUP and derive process motifs and structure motifs for steady-state covariances and correlations of node pairs in the mOUP. We discuss similarities and differences between the mechanisms for these covariances and correlations. In \cref{sec:conclusions}, we conclude and discuss possible applications of our process-based approach to the study of motifs in networks. We also explain why the distinction between process motifs and structure motifs is important for many (but not all) dynamical systems on networks. We discuss a few technical points in the appendices.


\section{Process motifs and structure motifs}\label{sec:definitions}

In this section, we define process motifs and structure motifs. In \cref{sec:theory}, we give a brief introduction to relevant graph-theoretical concepts. In \cref{sec:concept_defs}, we introduce walk graphs. We then define process motifs as weakly connected walk graphs and structure motifs as connected graphs. In \cref{sec:matching}, we introduce the concepts of matching process motifs and matching structure motifs. (These concepts are useful for our calculations in \cref{sec:example}.) To further illustrate the conceptual difference between process motifs and structure motifs, we compare methods for counting occurrences of process motifs and structure motifs in \cref{sec:motif_counts}. In \cref{sec:litreview}, we review prior uses of process motifs and structure motifs in the study of networks.


\subsection{Some graph-theoretical concepts}\label{sec:theory} 

We now give definitions for \textit{walks} and \textit{trails} on networks and \textit{paths} in networks. These words and other terminology for graph-theoretical concepts are often used ambiguously, and we will need to distinguish these concepts clearly for our work in the present paper.

We consider a \textit{graph} to be an ordered tuple $(V,E)$ that consists of a set $V$ of nodes and a set $E\subseteq V\times V$ of edges \cite{Bollobas2013}. Graphs can have self-edges, in which a node is connected to itself via an edge. They cannot have multi-edges. If the graph is \textit{directed}, its edges $e\in E$ are ordered pairs of nodes. If the graph is undirected, its edges $e\in E$ are unordered pairs of nodes. A \textit{weighted} graph is an ordered tuple $(V,E,W)$; it has a node set $V$ and an edge set $E$ as before, and there is also a map $W$ that assigns a \textit{weight} to each edge in $E$. For the remainder of the present paper, we exclude $W$ from our notation for graphs. However, our definitions and results hold for both weighted and unweighted graphs, and we assume that edges can have weights.

A \textit{subgraph} $(V',E')$ of a graph $(V,E)$ is a graph that consists of a node set $V'\subseteq V$ and an edge set $E'\subseteq E$ \cite{Bollobas2013}. A \textit{supergraph} $(V'',E'')$ of a graph $(V,E)$ is a graph with node set $V''\supseteq V$ and an edge set $E''\supseteq E$ \cite{Trudeau2013}.

We distinguish between walks and trails on graphs and paths in graphs. Consider a directed or undirected graph $(V,E)$. A \textit{walk} in this graph is a sequence
\begin{align}
	\omega=(v_{i_1},e_{i_1,i_2},v_{i_2},e_{i_2,i_3}, \dots, e_{i_{\ell-1},i_\ell},v_{i_\ell}, e_{i_\ell, i_{\ell+1}},v_{i_{\ell+1}})\nonumber
\end{align}
of nodes $v_{i_1},v_{i_2},\dots,v_{i_\ell},v_{i_{\ell+1}}\in V$ and edges $e_{i_1,i_2},e_{i_2,i_3},\dots,e_{i_{\ell-1},i_\ell},e_{i_\ell, i_{\ell+1}}\in E$ such that each edge $e_{i,j}$ starts at node $v_i$ and ends at node $v_j$ \cite{Bollobas2013}. The number $\ell$ indicates the number of edges in a walk. We call $\ell$ the \textit{length} of the walk. If no edge in $E$ appears more than once in $\omega$, the walk $\omega$ is also a \textit{trail} \cite{Bollobas2013}. 
If no node in $V$ and no edge in $E$ appear more than once in $\omega$, one can use the set of nodes in $\omega$ and the set of edges in $\omega$ to construct a \textit{path}. 
A \textit{path} is a subgraph $(V',E')$ that consists of a node set $V' \subseteq V$ and an edge set $E' \subseteq E$ that one can combine to construct a sequence 
\begin{align}
	(v_{i_1},e_{i_1,i_2},v_{i_2},e_{i_2,i_3}, \dots, e_{i_{\ell-1},i_\ell},v_{i_\ell}, e_{i_\ell, i_{\ell+1}},v_{i_{\ell+1}})\nonumber
\end{align}
of nodes and edges \cite{Bollobas2013}. The number $\ell$ is the \textit{length} of the path.

A path is a subgraph of a graph. By contrast, a walk is a combination (with repetition allowed) of a graph's nodes and edges\footnote{Other researchers have defined a path to be a combination of nodes and edges without repetitions \cite{Newman2018}. Using this definition, a path is a special case of a walk. For our work, it is crucial to distinguish between paths and walks as two fundamentally different concepts, where the former is related to processes on networks and the latter is related to graph structure.}. One can use walks to describe many processes on graphs \cite{Alon2011, Godsil2011, Nguyen2018, Spielman2012}. Additionally, one can consider the sequence of nodes and edges in a walk to be the temporal sequence of nodes and edges that a signal, a person, or some other entity traverses.

A \textit{closed walk} of length $\ell$ is a sequence
\begin{align}
	w=(v_{i_1},e_{i_1i_2},v_{i_2},e_{i_2i_3},\dots, e_{i_{\ell-1}i_\ell},v_{i_\ell}, e_{i_\ell i_1},v_{i_1})\nonumber
\end{align}
of nodes and edges \cite{Bollobas2013}. A \textit{cycle} of length $\ell$ is a subgraph $(V',E')$ that consists of a node set $V'\subseteq V$ and an edge set $E'\subseteq E$ that one can combine to construct a closed walk \cite{Bollobas2013}. One can think of a cycle as a closed path.
We say that a graph is \textit{cyclic} if it is a cycle. It is \textit{acyclic} if it is not a cycle and none of its subgraphs is a cycle. 

An undirected graph $(V,E)$ is \textit{connected} if there exists a path from $i$ to $j$ for every unordered pair $(i,j)$ of nodes in $V$. A directed graph $(V,E)$ is \textit{strongly connected} if there exists a path from $i$ to $j$ for every ordered pair $(i,j)\in V\times V$. A directed graph is \textit{weakly connected} if its corresponding undirected graph is connected.

A graph has an associated \textit{adjacency matrix} ${\bf A}=(a_{ij})$. If the graph is unweighted, $a_{ij}\in\{0,1\}$, where $a_{ij}=1$ indicates that there is an edge from node $j$ to node $i$.\footnote{There are different conventions for encoding the directions of edges in the adjacency matrix of a directed graph. We use the convention from Ref.\,\cite{Newman2018}.} For a weighted graph, the non-zero elements of ${\bf A}$ are $a_{ij}=w(e)$, where $w(e)$ is the weight of the edge $e$ from node $j$ to node $i$.

A \textit{multigraph} is like a graph, except that the edge set is now a multi-set $E\subseteq V\times V$, so an ordered node pair $(i,j)\in V\times V$ can be connected by multiple edges. We do not allow such multi-edges in the graphs in our paper, and we use the name ``networks'' for our graphs. In \cref{sec:concept_defs}, we define walk graphs and process motifs as special types of multigraphs that are associated with a network's structure.


\subsection{Walk graphs, process motifs, and occurrences of process motifs}\label{sec:concept_defs}

We define a \textit{walk graph} to be a weighted and directed multigraph $(\tilde V, \tilde E,\ell)$. We think of edges in $\tilde E$ as walks on a network. The walk-graph edge weights $\ell$ indicate the lengths of walks. A \textit{process motif} is a weakly connected walk graph. An \textit{occurrence} of a walk graph or process motif on a network $(V,E)$ is a labeling of nodes and edges in $(\tilde V, \tilde E,\ell)$ such that each walk-graph node $\tilde v \in \tilde V$ corresponds to a node $v\in V$ and each walk-graph edge $\tilde e \in \tilde E$ corresponds to a walk in $(V,E)$ with length $\ell(\tilde e)$. The labeling of nodes does not need to be bijective. Every node $\tilde V$ must correspond to exactly one node in $V$, but different nodes in $\tilde V$ can correspond to the same node in $V$ (see, for example, \cref{fig:walk_graphs}\panel{D}--\panel{F}). We say that a process motif \emph{occurs} on a network $(V,E)$ if there is at least one occurrence of the process motif on $(V,E)$.

When characterizing walk graphs, a useful property is the walk graph's \textit{(spatial) length} 
\begin{align}
	L:=\sum_{e\in \tilde E}\ell(e)\,.\nonumber
\end{align}
For the rest of our paper, we use the term ``length'' for a walk graph's spatial length\footnote{
One can also use a walk graph's 
\textit{duration} (i.e., \textit{temporal length}) to characterize it. The walk graphs that we derive in \cref{sec:example} are compositions of walks that start at their respective source nodes at the same time. The derivation motivates our definition of a walk graph's duration as $T:=\max_{e\in \tilde E}\ell(e)$. Walk-graph durations are not important for the derivations that we present in the present paper. However, they may be relevant for process motifs in networked dynamical systems in which edges have associated time delays.}.

To give some examples of walk graphs, we recall the walk graphs in \cref{fig:walk_graphs}. 
A walk graph that consists of a single edge corresponds to a single walk on the associated network (see, e.g., \cref{fig:walk_graphs}\panel{B}, \panel{D}, and \panel{G}). If a walk graph consists of a single self-edge, then the walk graph corresponds to a closed walk in the associated network (see, e.g., \cref{fig:walk_graphs} \panel{C}).

We noted in \cref{sec:theory} that one can interpret a walk to describe a type of process. 
One can thus use a walk graph to describe a composite process that consists of several walks. 
This interpretation motivates our definition of process motifs as weakly connected walk graphs.  
We consider a \textit{structure motif} to be a weakly connected graph. We consider an occurrence of a structure motif in a network $(V,E)$ to be a subgraph of $(V,E)$ that is isomorphic to the structure motif. 


\subsection{Matching process motifs and matching structure motifs}\label{sec:matching}

Consider the set $P_s$ of process motifs that occur on a structure motif $s$ and the set $S_p$ of structure motifs on which a process motif $p$ occurs.
If one does not specify a number $|\tilde V|$ of nodes and a length $L$ of a process motif, the set $P_s$ for any $s$ with one or more edges includes infinitely many process motifs because a process motif can use each edge of the structure motif infinitely many times. Conversely, for a given process motif $p$, the set $S_p$ includes infinitely many structure motifs because one can add nodes or edges to any $s\in S_p$ to obtain another element of $S_p$.

Most elements in $P_s$ are very long process motifs, and most elements of $S_p$ are very large structure motifs. Traditionally, studies of motifs in networks have focused on small motifs: process motifs with length $L\leq4$ \cite{Barnett2011, Lizier2012, Novelli2020} and structure motifs with up to five nodes \cite{Milo2002, Yaveroglu2014}. To associate small process motifs with small structure motifs and vice versa, we define \textit{matching process motifs} and \textit{matching structure motifs}. For a given process motif $p$, a matching structure motif $s^*_p$ is a structure motif on which $p$ occurs while using each edge in $s^*_p$ exactly once. Conversely, for a given structure motif $s$, a \textit{matching process motif} is a process motif that occurs on $s$ while using each edge in $s$ exactly once.

For a structure motif $s$ with a finite number of edges, the set $P^*_s$ of matching process motifs has a finite number of elements. For a process motif $p$ with a finite length $L$, the set $S^*_p$ of matching structure motifs 
has a finite number of elements.

\begin{figure*}[t]
\centering
\includegraphics[trim={3.1cm 1.25cm 2.4cm 1.25cm}, 
clip,width=\textwidth]{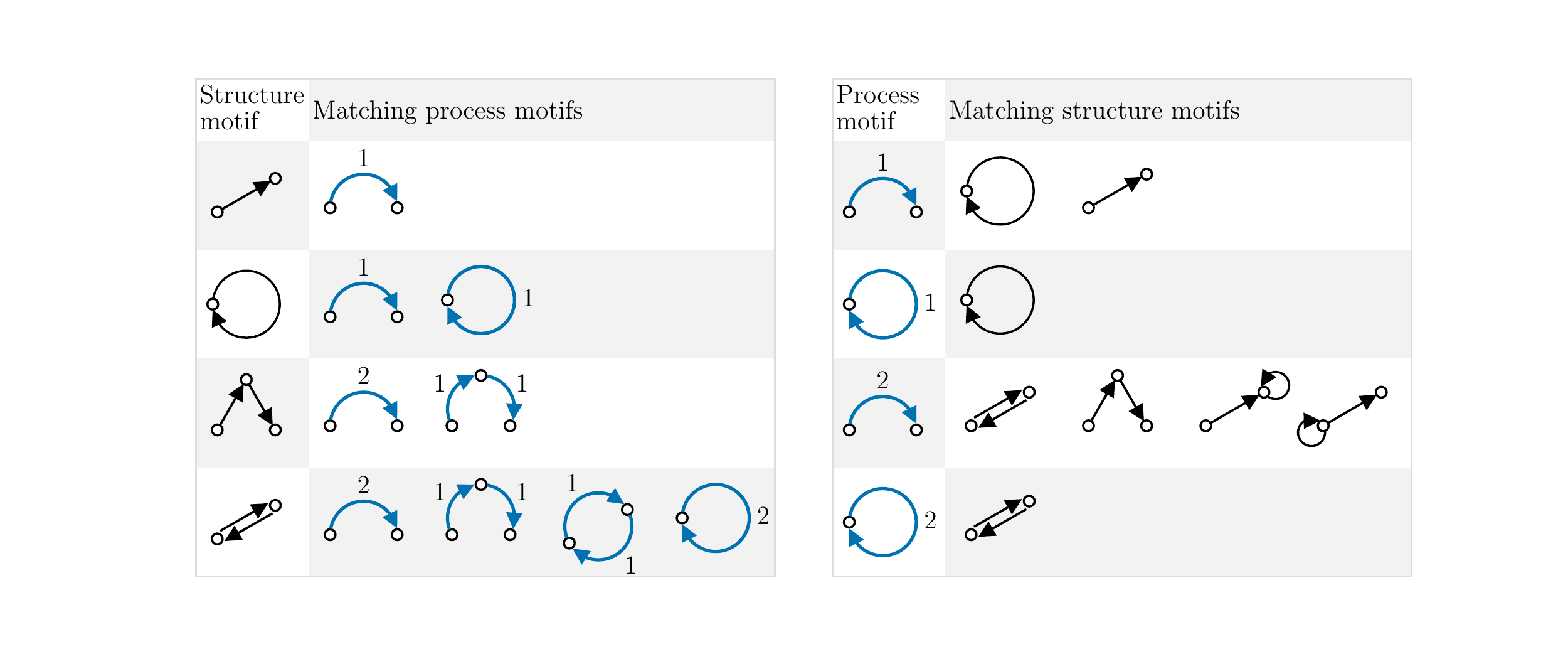}
\caption[Matching sets of structure motifs and matching sets of process motifs.]{Matching sets of structure motifs and matching sets of process motifs. On the left, we show four structure motifs and their sets of matching process motifs. On the right, we show four process motifs and their sets of matching structure motifs.}
\label{fig:matching}
\end{figure*}

In \cref{fig:matching}, we show sets of matching process motifs and sets of matching structure motifs for several structure motifs and process motifs, respectively. Structure motifs that do not include cycles have only acyclic matching process motifs.
Therefore, for a given number of edges, structure motifs that include cycles (e.g., the structure motifs in the second and fourth rows of the left table of \cref{fig:matching}) have more matching process motifs than acyclic structure motifs (e.g., the structure motifs in the first and third rows of the left table of \cref{fig:matching}). Accordingly, acyclic process motifs have more matching structure motifs than cyclic process motifs.

In general, a structure motif can have many matching process motifs and a process motif can have many matching structure motifs. Motif-based research that aims to link network structure to dynamics on networks requires careful consideration of these matching motifs. In \cref{sec:example}, we demonstrate the importance of these considerations using steady-state covariance and steady-state correlation of a multivariate Ornstein--Uhlenbeck process (mOUP) as an example.


\subsection{Counts of process motifs and structure motifs}\label{sec:motif_counts}

It is common for studies of motifs to associate motifs with a ``count'', ``number'', or ``frequency'' to indicate the prevalence of occurrences of a given motif in a given system \cite{Milo2002, Wu2012, Yaveroglu2014}. The \textit{count} (i.e., number) of a structure motif $s$ in a network $(V,E)$ is the number of occurrences of $s$ in $(V,E)$ (i.e., the number of labeled subgraphs of $(V,E)$ that are isomorphic to $s$). We consider the count of a process motif $p$ in an unweighted network $(V,E)$ to be the number of occurrences of $p$ in $(V,E)$. 

For a weighted network $(V,E,W)$, it is useful to weight each occurrence of a process motif by the product $\pi:=\prod_e w(e)$, where one takes the product of the weights of edges $e\in E$ that the walks in $p$ traverse. (If the walks in $p$ traverse an edge $k$ times, the corresponding edge weight $w(e)$ appears in $\pi$ with multiplicity $k$.) For weighted networks, we define the count of $p$ to be to the sum of edge-weight products $\pi$ for each occurrence of $p$ on $(V,E,W)$.

\begin{figure*}[t]
\centering
\includegraphics[trim={3.4cm 1.8cm 2.6cm 1.5cm},
clip,width=1\textwidth]{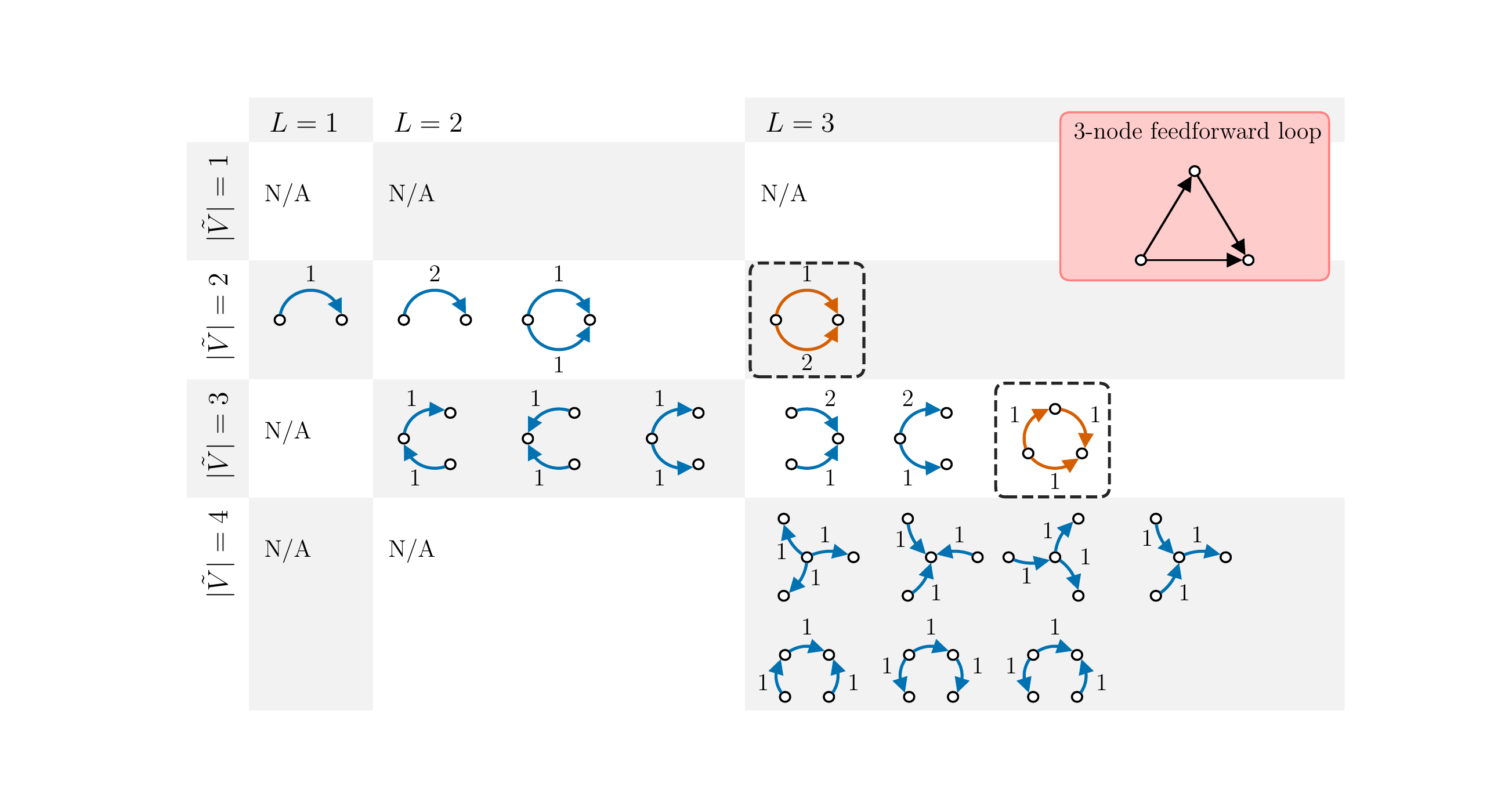}
\caption[Process motifs on a 3-node feedforward loop.]{Process motifs on a 3-node feedforward loop. We sort process motifs according to their length $L$ and their number $|\tilde V|$ of nodes. The numerical edge labels on process motifs indicate the length of an edge. Some process motifs occur on the 3-node feedforward loop but not on the 3-node feedback loop. We use orange edges and boxes with dashed boundaries to distinguish these process motifs from others. The pink inset in the top-right corner shows the structure of a 3-node feedforward loop.
}
\label{fig:structureFF}
\end{figure*}

\begin{figure*}[t]
\centering
\includegraphics[trim={3.4cm 1.8cm 2.6cm 1.5cm},
clip,width=1\textwidth]{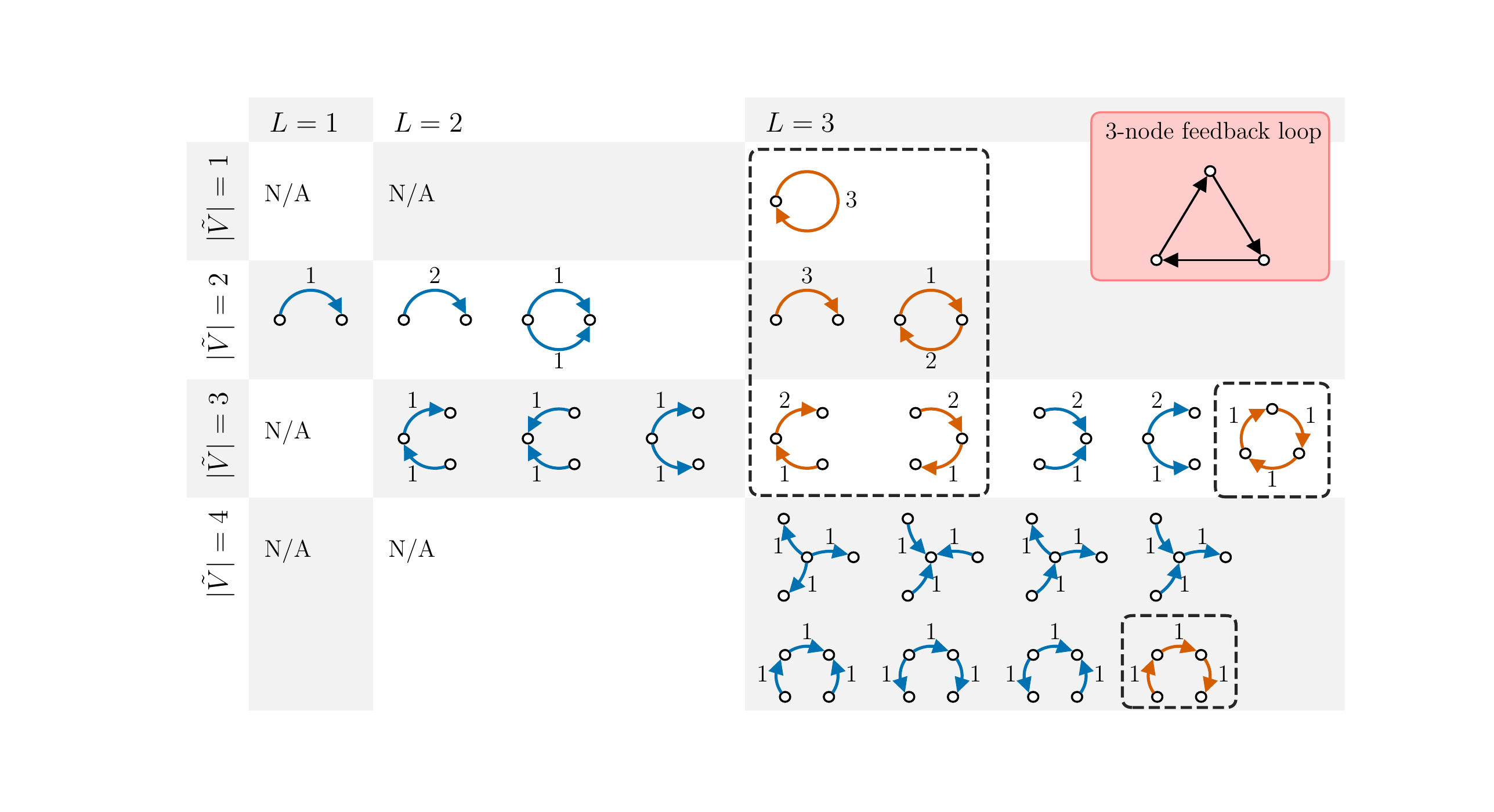}
\caption[Process motifs on a 3-node feedback loop.]{Process motifs on a 3-node feedback loop. We use orange edges and boxes with dashed boundaries to distinguish process motifs that occur on the 3-node feedback loop but not on the 3-node feedforward loop. 
}
\label{fig:structureFB}
\end{figure*}

The counts of structure motifs in a network and the counts of process motifs are related to each other. Each structure motif $s$ has an associated set $P_s$ of process motifs that occur on it. Consequently, a change in the number of occurrences of $s$ (i.e., the count of $s$) in a graph $(V,E)$ leads to a change in the counts of each $p\in P_s$ on $(V,E)$. To illustrate this relationship between counts of structure motifs and counts of process motifs, we consider two small example networks: a 3-node feedforward loop and a 3-node feedback loop \cite{Milo2002}. 
In \cref{fig:structureFF}, we show all length-$L$ walk graphs that occur on a 3-node feedforward loop for $L\leq3$. In \cref{fig:structureFB}, we show all length-$L$ walk graphs that occur on a 3-node feedback loop for $L\leq3$.

By comparing \cref{fig:structureFF,fig:structureFB}, we observe that some process motifs occur on the 3-node feedforward loop but not on the 3-node feedback loop, and vice versa. The differences between the process motifs on the 3-node feedforward loop and those on the 3-node feedback loop illustrate that the structure of a network constrains the structures of process motifs that occur on it. The 3-node feedforward loop is an acyclic network with a maximum trail length of 2. Because the feedforward loop is a directed acyclic graph (DAG), its associated walk graphs are also acyclic. Walk graphs on a DAG with a maximum trail length of 2 cannot have edges of length $\ell(e)>2$. 

The structure of the 3-node feedback loop leads to other constraints on the structures of associated process motifs. For example, a process motif that occurs on the 3-node feedforward loop but not on the 3-node feedback loop is the circular process motif with $|\tilde V|=2$ and $L=3$ in \cref{fig:structureFF}. This process motif consists of a length-1 edge and a length-2 edge that share both their starting node and their ending node.


\subsection{Previous work on process motifs and structure motifs} \label{sec:litreview}

To the best of our knowledge, previous research on network motifs has not distinguished explicitly between process motifs and structure motifs. Instead, studies have been concerned either with process motifs or with structure motifs, and they have used used the term ``network motifs'' for either of them. In this section, we give an overview of research on ``network motifs'' and explain which of the reviewed studies concern process motifs and which concern structure motifs.

Many reviews of network motifs have credited Milo et al.~\cite{Milo2002} for the idea of characterizing networks by connected subgraphs that are more frequent in a network than one would expect \cite{Alon2007, Ciriello2008, Masoudi-Nejad2012}. (The expectation is usually based on the frequency of connected subgraphs in a configuration model \cite{Milo2002, Schlauch2015, Stone2019}.) Other researchers have indicated that the search for frequent patterns in networks was already a topic of interest in, for example, ecology in the 1970s \cite{Stone2019}.

Milo et al.~\cite{Milo2002} compared several gene-regulatory networks, a neural network of the worm \textit{C.\,elegans}, several food webs, several electronic circuits, and the World Wide Web. They viewed gene-regulatory networks and neural networks as systems ``that perform information processing'' and reported that these networks have similar overrepresented connected subgraphs. They also reported that other networks, such as food webs and the World Wide Web, do not have similar overrepresented subgraphs as the considered gene-regulatory networks and the \textit{C.\,elegans} neural network. Subsequently, many researchers have studied various networks by identifying overrepresented connected subgraphs (e.g., see \cite{Conant2003, Hong-Lin2014, Ohnishi2010, Sporns2004, Takes2018}). In the corresponding publications, researchers used ``motif'' or ``network motif'' to refer to an overrepresented connected subgraph, which is a structure motifs or an occurrence of a structure motifs.

Closely related to the idea of characterizing networks by examining overrepresented subgraphs is the idea of characterizing networks based on the numbers or frequencies of one or several specified subgraphs \cite{Battiston2017, Bianconi2004, Dechery2018, Garcia2012, Gleiss2001, Iturria-Medina2008, Maayan2008, Manrubia2003, Shen2012, Sporns2000, Sporns2000a, Vazquez2004, Zhigulin2004}. For example, several researchers have used the number of triangles in an undirected network's structure to characterize networks \cite{Battiston2017, Dechery2018} or to explain aspects of dynamics on these networks \cite{Garcia2012, Zhigulin2004}. Others have used the numbers of different structure motifs with three or four nodes to compare networks \cite{Bianconi2004, Gleiss2001, Iturria-Medina2008, Manrubia2003, Shen2012, Vazquez2004} or to explain aspects of dynamics on them \cite{Maayan2008, Sporns2000, Sporns2000a}.
In some of these studies, researchers have considered ``network motifs'' to be connected subgraphs without the requirement of overrepresentation with respect to a null model \cite{Garcia2012, Manrubia2003}. The ``network motifs'' in these studies are also structure motifs or occurrences of structure motifs.

Estrada and Rodr\'iguez-Vel\'azquez~\cite{Estrada2005} proposed a measure of centrality that exploits the relationship between structure motifs and process motifs in a network. Their centrality measure, which is called ``subgraph centrality'', is a weighted sum of closed walks that start and end at a node. Noting that ``each closed walk is associated with a connected  subgraph'' \cite{Estrada2005}, Estrada and Rodr\'iguez-Vel\'azquez concluded that one can use a weighted sum of closed walks that start and end at a node as a measure of the count of cyclic graphlets that include that node. Their rationale for proposing subgraph centrality thus makes implicit use of the fact that each process motif that consists of a single closed walk has a corresponding matching structure motif that is a cycle.

In several theoretical studies of dynamical systems on networks, researchers have used process motifs when interpreting the results of their derivations \cite{Barnett2009, Barnett2011, Hu2014, Jovanovic2016, Lizier2012, Novelli2020, Pernice2011, Trousdale2012}. In theoretical neuroscience, a common approach to connect network structure with system functions is to linearize a nonlinear dynamical system about an equilibrium point and consider the effect of small perturbations on the dynamics. The strength of the coupling between a system's nodes affects how perturbations change the evolution of a system state. In a weakly coupled system with a parameter $\epsilon$ that tunes the coupling strength, one can sometimes approximate the effect of a perturbation on a system state by expanding the time evolution of the perturbed system in terms of increasing order in $\epsilon$ and truncating the resulting expression at some order of $\epsilon$. Researchers have used this approach to find process motifs for ``neural complexity'' \cite{Barnett2009, Barnett2011, Tononi1994}, information content \cite{Lizier2012}, transfer entropy \cite{Novelli2020}, cross-correlations \cite{Pernice2011, Trousdale2012}, and other properties of stochastic dynamical systems on networks \cite{Hu2014, Jovanovic2016}. In these studies, the order of $\epsilon$ in the approximation indicates the length or duration of the corresponding process motif.

Barnett et al.~\cite{Barnett2009, Barnett2011} considered the mOUP on a network and derived an approximation for neural complexity up to third order in $\epsilon$. They associated the terms of their approximation with ``graph motifs'' with up to three edges \cite{Barnett2011}. These graph motifs are process motifs with length $L\leq3$. Lizier et al.~\cite{Lizier2012} derived an approximation for the information content of a multivariate Gaussian autoregressive process on a network to fourth order in $\epsilon$ and associated terms of the approximation with process motifs with up to four edges. For the same dynamical system, Novelli et al.~\cite{Novelli2020} recently derived process motifs for pairwise transfer entropy to fourth order in $\epsilon$. Pernice et al.~\cite{Pernice2011} derived an approximation for the mean covariance of spiking rates in a system of coupled Hawkes processes \cite{Hawkes1971} to arbitrary order in $\epsilon$.
Trousdale et al.~ \cite{Trousdale2012} derived an approximation for individual cross-correlations of a system of coupled integrate-and-fire neurons \cite{Boergers2017} to arbitrary order in $\epsilon$. They associated each order of their approximation with a ``submotif'' that includes time-ordered edges. 
These submotifs are unions of process motifs. Hu et al.~\cite{Hu2014} approximated a measure of ``global coherence'' for the mOUP on a network to arbitrary order in $\epsilon$. They associated each order of their approximation with a normalized count (which they called a ``motif cumulant'') of a so-called ``$(n,m)$ motif''. The ``$(n,m)$ motifs'' are equivalent to the process motifs for mean covariance of spiking rates in a Hawkes process \cite{Pernice2011} and to the process motifs that we derive for covariances of the mOUP in \cref{sec:example}. Jovanovic and Rotter~ \cite{Jovanovic2016} derived approximations for covariance and the third joint cumulant, which is a measure of dependence between three variables, for a network of coupled Hawkes processes. They associated their approximation of covariance with 2-edge process motifs and their approximation of the third joint cumulant with process motifs with three or more edges.

Other types of dynamical systems on networks that are relevant to the perspective of the present paper include the spread of opinions \cite{Lehman2018} and the spread of infectious diseases \cite{Pastor-Satorras2015}. In probabilistic compartment models on networks, which are the most common type of model for studying infectious diseases on networks, the probability that a node is infected can depend on the infection probability of other nodes \cite{Kiss2017}. For a subset of the nodes, it is common to approximate joint moments of infection probabilities by products of moments (if there is only a single node in the subset) or joint moments (if there are two or more nodes in the subset) of the node(s) \cite{Demirel2014, Kiss2017}. When making such an approximation, one selects the joint moments of node-infection probabilities on some motifs --- typically, connected pairs or connected triples of nodes --- to be relevant for a spreading process and other joint moments to be negligible \cite{Chandra2020, House2009}. The motifs in these models can be process motifs or structure motifs. 
Researchers have used DAGs to describe the spread of behavior, norms, and ideas \cite{Oh2018} and the spread of infectious diseases \cite{Haydon2003, Juul2020} on networks. One can view subgraphs of these so-called ``dissemination trees'' \cite{Oh2018} and ``epidemic trees'' \cite{Haydon2003, Juul2020} as process motifs.

For many studies of the spread of infectious diseases, either the choice of compartment model (e.g., susceptible--infected--recovered \cite{Newman2002, Radicchi2016}) or the choice of network structure (e.g., if it is locally tree-like \cite{Bianconi2017, Larremore2012}) constrains the number of relevant process motifs such that some structure motifs have only one relevant process motif that occurs on it. Because of this one-to-one correspondence between process motifs and structure motifs, the distinction between them is irrelevant for these models of disease spread, provided that one considers only structure motifs with one corresponding process motif. In \cref{sec:matter}, we discuss when the distinction between process motifs and structure motifs is relevant and when it is not.


\section{Using process and structure motifs to study functions of dynamics on networks}\label{sec:part2}

In this section, we motivate the use of process motifs for the study of dynamical systems on networks. We formally define contributions of occurrences of process motifs and structure motifs to real-valued functions of the state of a dynamical system on a network. We focus on linear dynamical systems. In general, one cannot use the same approach to directly study nonlinear dynamical systems, although one can apply our approach to linearizations of them.


\subsection{Linking process motifs to properties of dynamics on networks}\label{sec:links}

Consider a linear dynamical system
\begin{align}
	\frac{d{\bf x}_t}{dt} = {\bf F}({\bf A}){\bf x}_t\,,\label{eq:linsys}
\end{align}
where ${\bf x}_t$ is a column vector that describes the current system state and ${\bf F}$ is a matrix-valued function of the adjacency matrix ${\bf A}$ of a network. The system has the initial state ${\bf x}_{0}={\bf x}_{t=0}$. Observables of the linear dynamical system in \eq{linsys} are functions of ${\bf x}_t$ and ${\bf F}$, and they are thus functions of ${\bf A}$ and ${\bf x}_0$. (For systems at steady state or a system with identical initial values ${x_0}_1={x_0}_2={x_0}_3=\dots$,
 one can often remove the dependence on ${\bf x}_0$ and describe functions of the dynamical system as functions of only ${\bf A}$.) One can thus view a function of the linear dynamical system \eqref{eq:linsys} as a superposition of walks or a superposition of process motifs on a network.

This view motivates the approach that we take in the present paper. We study how a function of a linear dynamical system emerges via the superposition of process motifs, which are structured sets of walks that occur on an associated network. After identifying relevant process motifs for a given property of a dynamical system on a network, one can establish links between dynamics on networks and network structure by identifying the structure motifs on which the relevant process motifs occur. This approach results in (1) a set of structure motifs that contribute to the desired system function and (2) a discovery of the mechanisms by which these structure motifs contribute to this function. When it is possible to quantify the contribution of process motifs to a function of interest, one can also quantify the contribution of structure motifs. In \cref{sec:example}, we demonstrate our approach using the covariances and the correlations in the mOUP at steady state. In the remainder of \cref{sec:part2}, we explain how we formalize links between process motifs and structure motifs.


\subsection{Contributions of occurrences of process motifs and structure motifs}\label{sec:contributions}

We now discuss how we characterize the importance of motifs to a system property $Y$ via contributions of their occurrences. We first discuss two conceptually different notions of the contribution of a motif to a system property. We then explain how one can express $Y$ as a weighted sum of counts process motifs and as a weighted sum of counts of structure motifs.


\subsubsection{Contributions of motifs}

\begin{figure*}[t]
\centering
\includegraphics[trim={2.05cm 1.7cm 1.8cm 0.8cm},
clip,width=0.7\textwidth]{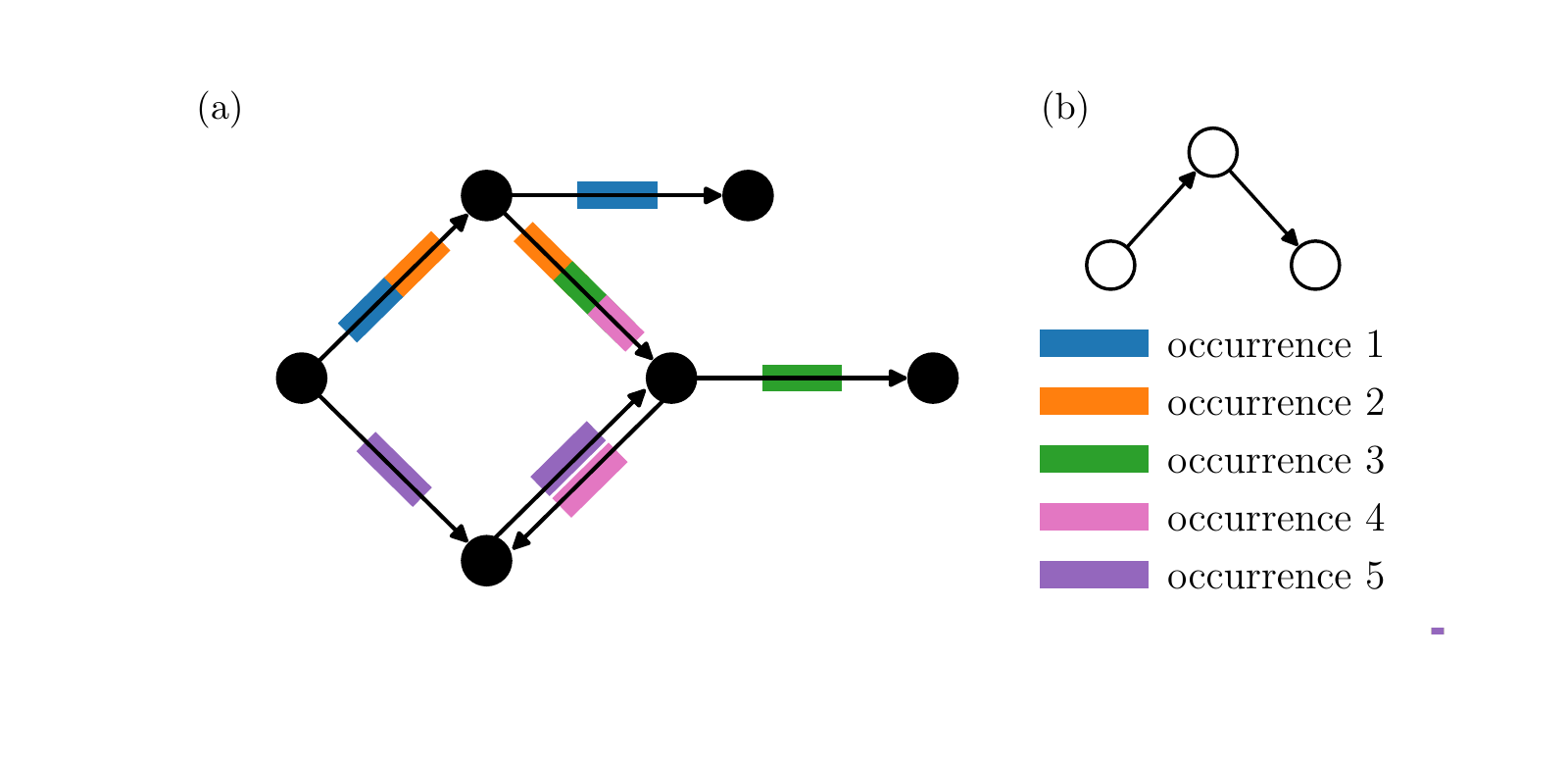}
\caption{Occurrences of structure motifs. In \panel{A}, we show an example network. In \panel{B}, we show a structure motif, which is a length-2 path. This structure motif has five occurrences on the example network. We use colored rectangles to indicate the edge sets of the structure-motif occurrences.
}
\label{fig:occurrence_count}
\end{figure*}

Consider a network $(V,E)$ and a small structure motif $s$. In \cref{fig:occurrence_count}, we show an example network $(V,E)$ and use the length-2 path as an example of a small structure motif. In this example, there are five occurrences of $s$ in $(V,E)$. Consider a scalar property $Y=f(\bf{A})$ of a linear dynamical system on $(V,E)$, where ${\bf A}$ is the adjacency matrix of $(V,E)$, and suppose that we have derived that every length-2 path in $(V,E)$ has a contribution $c=2$ to $Y$. Should we say that the contribution of the structure motif $s$ to $Y$ is $2$ because that is the contribution of \textit{each} occurrence of $s$, or should we say that the contribution of $s$ is $10$ because the sum of contributions of \textit{all} occurrences of $s$ is equal to $10$? To distinguish between these two notions of motif contributions, we refer to the contribution of each occurrence of a process motif or structure motif as the motif's \textit{contribution per occurrence}, which we shorten to ``o-contribution''. We refer to the sum of contributions of all occurrences of a process motif or structure motif in a network as the motif's \textit{contribution per network}, which we shorten to ``n-contribution''. The values of o-contributions depend both on the dynamical system and on the system property $Y$. The values of n-contributions depend not only on the dynamical system and the system property $Y$, but also on the network $(V,E)$. We denote the o-contribution of a structure motif $s$ by $c_s$ and the n-contribution of that structure motif in a network $(V,E)$by $C_s$. The two types of contributions are related by the equation
\begin{align}
	C_s=n_{s}c_s\,, \nonumber
\end{align}
where $n_{s}$ is the count of $s$ in $(V,E)$. Similarly, the o-contribution $b_p$ of a process motif $p$ is related to the n-contribution $B_p$ of $p$ on $(V,E)$ by the equation 
\begin{align}
	B_p=n_pb_p\,,
\end{align}
where $n_p$ is the count of $p$ on $(V,E)$. 

In the present paper, we focus on o-contributions of process motifs and structure motifs. Our results are thus independent of network structure. For convenience, we will refer to o-contributions simply as ``contributions'' for the remainder of our paper.


\subsubsection{Contributions of process motifs and structure motifs}

Consider a scalar property $Y=f(\bf{A})$ of a linear dynamical system on a network with adjacency matrix ${\bf A}$. Assume that we have identified the relevant process motifs $p_1, p_2, \dots, p_k$ and the real-valued contributions $b_{p_1}, b_{p_2}, \dots, b_{p_k}$ to $Y$. In a network on which $p_i$ has the count $n_{p_i}$, we can compute $Y$ from 
\begin{align}
	Y = \sum_{i=1}^k b_{p_i}n_{p_i} \,,
\end{align}
which is a weighted sum of the counts of process motifs.

There are several ways that one can define the contribution of a structure motif to $Y$. For example, one can define the contribution $c_s$ of a structure motif $s$ to be the real-valued sum
\begin{align}\label{eq:total_c}
	c_s :=\sum_{p\in P_s}b_p
\end{align}
of all contributions $b_{p}$ over the set $P_s$ of all process motifs that occur on $s$. This association is intuitive and tends to be computationally easy. For a linear dynamical system, one can compute $c_s$ directly from $c_s=f({\bf A}')$, where ${\bf A}'$ is the adjacency matrix of the structure motif $s$. We refer to $c_s$ as the \textit{total contribution} of a structure motif to $Y$. 

There are some disadvantages of using $c_s$ to characterize the importance of structure motifs to $Y$. As we discussed in \cref{sec:matching}, any process motif $p$ occurs on many different structure motifs. (In fact, the set $S_p$ of structure motifs on which $p$ occurs is infinite.) Therefore, one cannot express $Y$ as the sum $\sum_{i} c_{s_i}n_{s_i}$ of structure-motif counts in a network, because this sum tends to overcount the contributions of many process motifs. 

Another disadvantage of using $c_s$ to measure the importance of a structure motif for a system property $Y$ is that total contributions are hard to interpret. If all $b_{p_i}>0$, large structure motifs tend to contribute much more to $Y$ than small structure motifs, because more process motifs occur on large structure motifs than on small structure motifs. Moreover, the total contribution $c_s$ of a structure motif $s$ depends strongly on the total contributions $c_s'$ of subgraphs $s'$ of $s$, because all process motifs that occur on any $s'$ can also occur on $s$. (For example, we demonstrate in \cref{sec:cov_contributions} that when $Y$ is either the steady-state covariance or steady-state correlation of a pair of nodes in the mOUP, the total contributions $c_s$ of a structure motif with $m$ edges has a very large positive correlation with the mean total contribution $\langle c_s'\rangle_{m-1}$ of subgraphs of $s$ with $m-1$ edges.) Consequently, a large total contribution $c_s$ does not necessarily indicate that $s$ is important for $Y$. It can instead indicate that $s$ is just a very large structure motif and/or that $s$ has subgraphs that are important for $Y$.

To address these two issues, we propose a different definition of the contribution of structure motifs to $Y$. The contribution $\specific_s$ of a structure motif $s$ is the sum of the contributions $b_{p}$ of process motifs $p$ that occur on $s$ but not on any subgraph of $s$. We refer to $\specific_s$ as the \textit{specific contribution} of a structure motif. One can express $Y$ as the sum
\begin{align}
	Y = \sum_{i} \specific_{s_i}n_{s_i} 
\end{align}
of weighted counts $n_{s_i}$ of structure motifs $s_i$. A contribution $\specific_s$ of a structure motif $s$ is not necessarily larger than the contribution $\specific_{s'}$ of a subgraph ${s'}$ of $s$. As we demonstrate in \cref{sec:example}, the specific contribution $\specific_s$ tends to be smaller than the specific contributions $\specific_{s'}$. A drawback of using specific contributions to characterize the importance of structure motifs to $Y$ is that specific contributions are much harder to compute than total contributions. 
One can compute the specific contribution of a structure motif $s$ recursively using the equation
\begin{align}
	\specific_s = c_s-\sum_{s'\subset s} \specific_{s'}\,,\label{eq:recurs}
\end{align}
where we use $s'\subset s$ to denote that $s'$ is a proper subgraph of $s$. Alternatively, one can use the mean total contributions $\langle c\rangle_{m'}$ (where $\langle\cdot\rangle_{m'}$ denotes the mean over all structure motifs with $m'$ edges) of subgraphs of $s$ with $m'$ edges to compute $\specific_s$. That is,
\begin{align}
	\specific_s =\sum_{m'=1}^{m}\left\{\left[\binom{m}{m'}\sum_{{\mathbf q}\in\mathcal Q({m-m'})}(-1)^{|{\bf q}|}{\bf q}{\tilde !}\right]\langle c\rangle_{m'}\right\}\,,\label{eq:direct}
\end{align}
where $m$ is the number of edges in $s$, the set $\mathcal Q({m-m'})$ is the set of integer compositions\footnote{An ``integer composition'' of a non-negative number $m$ is a sequence $(q_1,q_2,\dots,q_k)$ of positive integers, where $\sum_{k'=1}^{k}q_{k'}=m$ \cite{Eger2013}.} of ${m-m'}$, and the sequence $\mathbf q=(q_1, q_2, \dots, q_k)$ is an integer composition of ${m-m'}$ with $\indB$ elements. In \eq{direct}, we denote the number of elements in a sequence $\bf q$ by $|{\bf q}|$ and the multinomial coefficient of a sequence $\bf q$ of integers by ``${\bf q}\tilde{!}$''. We derive \eq{direct} in Appendix \ref{sec:app:direct}. For structure motifs with $m>2$ edges, it is computationally easier to calculate $\specific_s$ from \eq{direct} than from \eq{recurs}.


\section{Covariance and correlation for the multivariate Ornstein--Uhlenbeck process}\label{sec:example}

In this section, we demonstrate our process-based approach for studying motifs in networks. As an example, we examine steady-state covariances and steady-state correlations in the mOUP. We derive contributions of process-motif occurrences and the total and specific contributions of structure-motif occurrences to steady-state covariances and steady-state correlations in the mOUP. We then discuss the relationship between the specific contributions of structure motifs and network mechanisms that contribute to steady-state covariances and steady-state correlations in the mOUP.


\subsection{The Ornstein--Uhlenbeck process}

Uhlenbeck and Ornstein~\cite{Uhlenbeck1930} proposed a stochastic process to describe Brownian motion under the influence of friction. The mOUP is a popular model for coupled noisy systems, including neuronal dynamics \cite{Barnett2009}, stock prices \cite{Liang2011}, and gene expression \cite{Rohlfs2013}. In these studies, the mOUP with $n$ variables describes the dynamics on a network with $n$ nodes, where the state of each node represents a neuron, the value of a stock, or a gene-expression level.

One can describe the mOUP using the stochastic differential equation
\begin{align}
	d{\bf x}_{t+dt} = \theta(\epsilon {\bf A}-{\bf I}){\bf x}_t\,dt+\varsigma\, dW_t\,,\label{eq:OU}
\end{align}
where the column vector ${\bf x}_t\in\mathbbm R^N$ describes the state of the process. The process has an adjacency matrix ${\bf A}$, which can be directed and/or weighted, and a multivariate Wiener process $W_t$. The \textit{reversion rate} $\theta>0$, the \textit{noise strength} $\varsigma^2$, and the \textit{coupling parameter} $\epsilon>0$ are parameters of the mOUP.

We consider a signal to be a (temporary) deviation of a node's state from its mean. The coupling parameter $\epsilon$ sets the rate at which a signal's amplitude increases or decreases when it is transmitted from one node to another. The parameter $\theta$ is the rate at which a signal's amplitude increases or decreases over time. It thus determines the expected speed at which a node's state reverts to its mean. Because of this connection between $\theta$ and the speed of signal decay in the mOUP, many researchers refer to $\theta$ as the \textit{reversion rate} \cite{Ogbogbo2018, Sanderson2016, Tsai2011}.

If all eigenvalues of $\epsilon{\bf A}-{\bf I}$ have negative real parts, the mOUP has a single stationary distribution. We then say that the mOUP is a process with \textit{signal decay} because, in this process, a signal's amplitude decreases with time. A sufficient condition for signal decay is $\rho(\epsilon{\bf A})<1$, where $\rho(\cdot)$ is the spectral radius. For any network with finite edge weights, the mOUP in \eq{OU} is a process with signal decay if we choose $\epsilon$ to be sufficiently small.

The mOUP with signal decay is a Markov process. Its stationary distribution is a multivariate normal distribution $\mathcal N(0,\covmat)$ that is centered at $\langle {\bf x} \rangle=0$ with covariance matrix $\covmat:=\langle{{\bf x}_t{\bf x}_t^T}\rangle$ \cite{Barnett2009}. The mOUP with signal decay has the steady-state covariance matrix
\begin{align}
	\covmat = \frac{\varsigma^2}{2\theta}\sum_{L=0}^\infty\sum_{\substack{\ell=0}}
^\infty 2^{-L} \binom{L}{\ell} (\epsilon {\bf A})^\ell (\epsilon {\bf A}^T)^{L-\ell}\,.\label{eq:sigma}
\end{align}
Barnett et al.~\cite{Barnett2011} derived \eq{sigma} for the mOUP with $\theta=\varsigma=1$. In Appendix \ref{app:cov_deriv}, we show that \eq{sigma} also holds for arbitrary choices of $\theta>0$ and $\varsigma>0$.

In the remainder of this section, we derive and compare process motifs and structure motifs for the covariance, variance, and correlation of the mOUP at steady state.


\subsection{Process motifs for covariance and correlation at steady state}

We now derive process motifs and process-motif contributions of steady-state covariances and steady-state correlations in the mOUP.


\subsubsection{Process motifs for steady-state covariance}\label{sec:process_motifs}

We introduce the shorthand notation 
\begin{align}
	b_{L,\ell}:= \frac{\varsigma^2\epsilon^L}{2^{L+1}\theta}\binom{L}{\ell}\label{eq:bcov}
\end{align}
 and 
\begin{align}
	{\bf N}_{L,\ell}:={\bf A}^\ell ({\bf A}^T)^{(L-\ell)}
\end{align}
 to write
\begin{align}
	\covmat=\sum_{L=0}^\infty\sum_{\ell=0}^Lb_{L,\ell} {\bf N}_{L,\ell}\,.\label{eq:sigma2}
\end{align}
The $(i,j)$-th element of ${\bf N}_{L,\ell}$ corresponds to a count $n_p$ of process motifs $p$ for the steady-state covariance between nodes $i$ and $j$. The matrix ${\bf N}_{L,\ell}$ is not necessarily symmetric. However, the $(i,j)$-th element of ${\bf N}_{L,\ell}$ is equal to the $(i,j)$-th element of ${\bf N}_{L,L-\ell}$. 

\begin{figure*}[t]
\centering
\includegraphics[trim={3.8cm 0.7cm 2.2cm 0.7cm},clip,width=1\textwidth]{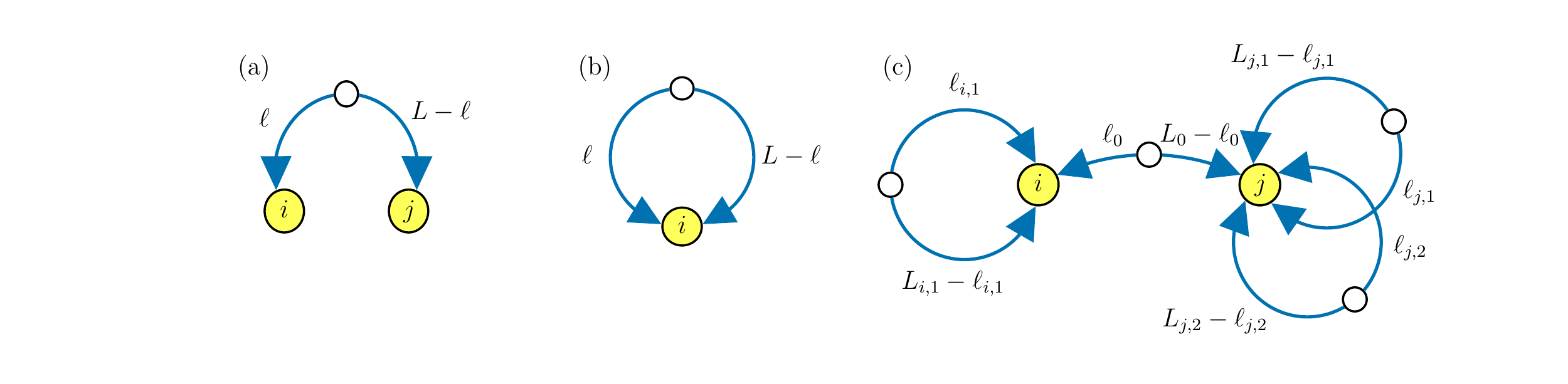} 
\caption{Process motifs for \panel{A} covariance, \panel{B} variance, and \panel{C} correlation at steady state.
}
\label{fig:covcorr}
\end{figure*}

Equation (\ref{eq:sigma2}) indicates that one can compute the covariances of the mOUP as a weighted sum of counts of process motifs. A process motif that contributes to the covariance between nodes $i$ and $j$ is a walk graph with three nodes and two edges. Two of the walk-graph nodes correspond to nodes $i$ and $j$ in the network. We refer to these walk-graph nodes as the \textit{focal nodes} of this process motif. All process motifs for covariance also include a third walk-graph node, which we call the ``source node'' and which can correspond to any node in a network.
Each edge in this process motif corresponds to a walk from the source node to one of the two focal nodes. We show a diagram of a process motif that contributes to covariance in \cref{fig:covcorr}\panel{A}. One can characterize a process motif of this form using the two parameters $L\in\{0,1,2,\dots\}$ and $\ell\in\{0,\dots,L\}$. The parameter $L$ is the length of a process motif, and the parameter $\ell$ is the length of the walk from the source node to node $i$. The contribution of each process motif to the covariance is $b_{L,\ell}$. It depends on the parameters $L$ and $\ell$ of the process motif and on the parameters $\epsilon$, $\varsigma$, and $\theta$ of the mOUP.

The process motifs for covariance are consistent with properties of covariation in a system of coupled random variables. A covariance $\covel_{ij}$ measures the joint ``variability'' of two random variables $x_i$ and $x_j$ \cite{Rice2006}, where we take variability to signify a variable's deviation from its mean. This joint variability of $x_i$ and $x_j$ can arise from several causes \cite{Reichenbach1956}:
\begin{enumerate}
\item Variability in $x_i$ induces variability in $x_j$ if there is a path from node $i$ to\linebreak node $j$.
\item Variability in $x_j$ induces variability in $x_i$ if there is a path from node $j$ to node $i$.
\item Variability in a third variable $x_k$ induces variability in both $x_i$ and $x_j$ if there are paths from $k$ to $i$ and from $k$ to $j$.
\end{enumerate}

\begin{figure*}[!b]
\centering
\includegraphics[trim={0.46cm 2.2cm 0.45cm 3.1cm},clip,width=0.6\textwidth]{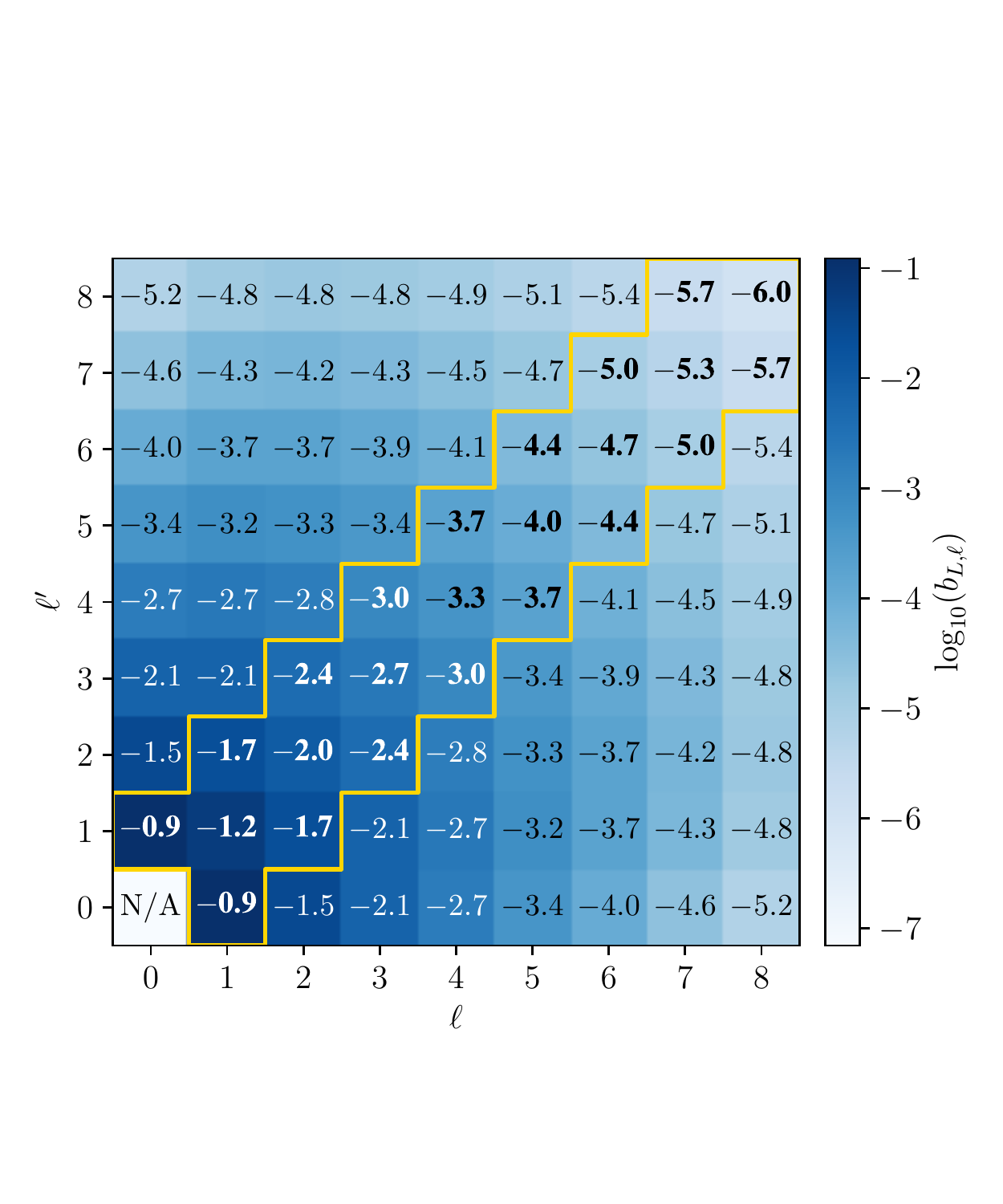}
\caption{Contributions $b_{L,\ell}$ of process motifs to the steady-state covariance with parameters $(L,\ell)$ for $\theta=1$, $\varsigma=1$, and $\epsilon=0.49$. The parameter $\ell$ specifies the length of one walk in the process motif. The length $\ell':=L-\ell$ is the length of the other walk in the process motif (see also \cref{fig:covcorr}(a)). The length $L$ increases along the diagonal from the bottom left to the top right. We indicate the parameter pairs for the largest contributions for each value of $L$ by bold labels and delineate them with yellow line segments.}
\label{fig:ccov}
\end{figure*}

We now compare the contributions of different process motifs to covariance. In \cref{fig:ccov}, we show the contributions $b_{L,\ell}$ of process motifs to steady-state covariance. The length $L$ increases along the diagonal from the bottom left to the top right. We indicate the parameter pairs with the largest contribution for each value of $L$ using bold labels and delineate them with yellow line segments. For even $L$, the contributions are maximal when $\ell=L/2$. For odd $L$, the contributions are maximal when $\ell=(L\pm1)/2$. Comparing the contributions of process motifs with different lengths, we find that short process motifs (see the bottom left) tend to contribute more to covariances than long process motifs. These results are consistent with the notion that covariances and correlations should decay with the distance that a signal travels \cite{Barzel2009}. The result that a process motif with $\ell = L/2$ contributes more to covariance than any other process motif with the same length $L$ is consistent with the notion that a signal that reaches two nodes $i$ and $j$ at the same time contributes more to the covariance or correlation between $i$ and $j$ than signals that reach $i$ and $j$ at different times.  


\subsubsection{Process motifs for steady-state variance}

A diagonal element of $\covmat$ indicates the variance of a node in the mOUP. By merging the focal nodes in \cref{fig:covcorr}\panel{A}, one obtains the process motifs that contribute to the steady-state variance of a node $i$ (see \cref{fig:covcorr}\panel{B}). Such a process motif includes two nodes and two edges. It includes a source node and a single focal node $i$. Its two edges correspond to two walks from the source node to node $i$.

We write
\begin{align}
	\covmat = \frac{\varsigma^2}{2\theta}{\bf I} + \covmat^{(1+)}\,,\label{eq:separation}
\end{align}
where 
\begin{align}
	\covmat^{(1+)}:=\sum_{L=1}^\infty\sum_{\ell=0}^Lb_{L,\ell} {\bf N}_{L,\ell}\,,\label{eq:covmat1}
\end{align}
to separate the intrinsic variance contribution $\frac{\varsigma^2}{2\theta}{\bf I}$ (which is independent of a network's structure) from structure-dependent variance contributions $\covmat^{(1+)}$ (which includes all terms of \eq{sigma2} that are $O(\epsilon^k)$ with $k\geq 1$). We interpret the two terms in \eq{separation} as indicators of two mechanisms by which variance arises in the mOUP:
\begin{enumerate}
\item Gaussian white noise in each node induces the $0$-th order contribution to variance. This effect contributes a value of $\varsigma^2/(2\theta)$ to the variance of the state variable $x_i$ at each node $i$. This contribution is determined by the noise strength $\varsigma$ and the reversion rate $\theta$. It is independent of a network's structure. 
\item The variance of a state variable $x_{i}$ exceeds its noise-induced base value of $\varsigma^2/(2\theta)$ when it receives input from other nodes via in-edges or from itself via a self-edge. For a node $i$, these network-dependent contributions are large when there are many occurrences of variance-increasing process motifs in which node $i$ is a focal node. This is the case when node $i$ is part of many cycles in a network or when many redundant paths or trails in a network connect other nodes to node $i$. Intuitively, cycles can reinforce the variance of the state of a node $i$. Redundant paths or trails that lead to node $i$ can amplify the input that $i$ receives from other nodes.
\end{enumerate}


\subsubsection{Process motifs for steady-state correlation}

The elements $\correl_{ij}$ of the correlation matrix $\corrmat$ are given by
\begin{align}
	\correl_{ij}:=\covel_{ij}/\sqrt{\covel_{ii}\covel_{jj}}\,.\label{eq:rij}
\end{align}
To replace the square root in the denominator of \eq{rij}, we use the Taylor-series expansion
\begin{align}
	\frac{1}{\sqrt{x}}=\sum_{k=0}^\infty \frac{(-1)^k}{2\cdot4^k}\binom{2k}{k}x_0^{-\frac{2k+1}{2}}(x-x_0)^k\,,\label{eq:expansion}
\end{align}
which we obtain from expanding about the point $x_0>0$. The radius of convergence of the expansion (\ref{eq:rij}) is $x_0$. We set $x_0=\varsigma^2/(2\theta)$ and substitute \eq{expansion} for $1/\sqrt{\covel_{ii}}$ and $1/\sqrt{\covel_{jj}}$ to obtain
\begin{align}
	\correl_{ij}&=
\frac{2\theta}{\varsigma^2}\covel_{ij}\sum_{\kOne=0}^\infty\sum_{\kTwo=0}^\infty\left(-\frac{\theta}{2\varsigma^2}\right)^{\kOne+\kTwo}\binom{2\kOne}{\kOne}\binom{2\kTwo}{\kTwo}
\left(\covel_{ii}-\frac{\varsigma^2}{2\theta}\right)^\kOne\left(\covel_{jj}-\frac{\varsigma^2}{2\theta}\right)^\kTwo\nonumber\\
	&=\frac{2\theta}{\varsigma^2}\covel_{ij}\sum_{\kOne=0}^\infty\sum_{\kTwo=0}^\infty\left(-\frac{\theta}{2\varsigma^2}\right)^{\kOne+\kTwo}\binom{2\kOne}{\kOne}\binom{2\kTwo}{\kTwo}{\left(\covel_{ii}^{(1+)}\right)}^\kOne{\left(\covel_{jj}^{(1+)}\right)}^\kTwo\,,\label{eq:correl1}
\end{align}
where $\covel_{ii}^{(1+)}$ and $\covel_{jj}^{(1+)}$ are elements of $\covmat^{(1+)}$ (see \eq{covmat1}). Equation (\ref{eq:correl1}) is a valid expression for $\correl_{ij}$ whenever the sums in \eq{correl1} converge. Whenever \eq{correl1} converges, we say that the mOUP has \textit{short-range} signal decay. A sufficient condition for short-range signal decay is $\|\epsilon {\bf A}\|_2<1/2$, where $\|\cdot\|_2$ denotes the Hilbert--Schmidt norm. When ${\bf A}$ is the adjacency matrix of a strongly connected network with non-negative edge weights, another sufficient condition for short-range signal decay is $\rho(\epsilon{\bf A})<1/2$. We derive these sufficient conditions for short-range signal decay in Appendix \ref{app:short_range}. 

From \eq{covmat1}, we see that one can express $\covel_{ii}^{(1+)}$ as a sum over the two indices $L$ and $\ell$. Consequently, one can express the $k$-th power of $\covel_{ii}^{(1+)}$ as a sum over the $2k$ indices $L_{1},\ell_{1},L_{2},\ell_{2},\dots,L_{k},\ell_{k}$. We use the multisets
\begin{align}
	\cvec_i &:=\{(L_{i,1},\ell_{i,1}),(L_{i,2},\ell_{i,2}),\dots,(L_{i,\kOne},\ell_{i,\kOne})\}\,,\nonumber\\
	\cvec_j &:=\{(L_{j,1},\ell_{j,1}),(L_{j,2},\ell_{j,2}),\dots,(L_{i,\kTwo},\ell_{i,\kTwo})\}\nonumber
\end{align}
of pairs of indices to write
\begin{align}
	\correl_{ij}=
\frac{2\theta}{\varsigma^2}\covel_{ij}\sum_{\cvec_i,\cvec_j}&\left\{
\left(-\frac{\theta}{2\varsigma^2}\right)^{|\cvec_i|+|\cvec_j|}\binom{2|\cvec_i|}{|\cvec_i|}\binom{2|\cvec_j|}{|\cvec_j|}\right.\nonumber\\
&
\left.\phantom{,}\times
\left[\prod_{\lastIndex=1}^{|\cvec_i|}b_{L_{i,\lastIndex},\ell_{i,\lastIndex}}({\bf N}_{L_{i,\lastIndex},\ell_{i,\lastIndex}})_{ii}\right]
\left[\prod_{\lastIndex=1}^{|\cvec_j|}b_{L_{j,\lastIndex},\ell_{j,\lastIndex}}({\bf N}_{L_{j,\lastIndex},\ell_{j,\lastIndex}})_{jj}\right]
\right\}\nonumber
\,,
\end{align}
where $|\cvec_i|$ denotes the number of pairs in $\cvec_i$. We use $\sum_{\cvec_i,\cvec_j}$ to denote the double summation over all possible multisets of pairs $(L,\ell)$ of non-negative integers with $\ell\leq L$.
We can thus express the steady-state correlation as a weighted sum of counts of process motifs:
\begin{align}
	\correl_{ij}=\sum_{L_0,\ell_0,\cvec_i,\cvec_j}b_{L_0,\ell_0,\cvec_i,\cvec_j}{\bf N}_{L_0,\ell_0,\cvec_i,\cvec_j}\,,\nonumber
\end{align}
where
\begin{align}
	b_{L_0,\ell_0,\cvec_i,\cvec_j}:=\frac{2\theta}{\varsigma^2}\sum_{\cvec_i,\cvec_j}&\left(-\frac{\theta}{2\varsigma^2}\right)^{|\cvec_i|+|\cvec_j|}\binom{2|\cvec_i|}{|\cvec_i|}\binom{2|\cvec_j|}{|\cvec_j|}
	 b_{L_0,\ell_0}\prod_{(L,\ell)\in\cvec_i}b_{L,\ell}\prod_{(L,\ell)\in\cvec_j}b_{L,\ell}\label{eq:bcorr}
\end{align}
and 
\begin{align}
	{\bf N}_{L_0,\ell_0,\cvec_i,\cvec_j}:= \left({\bf N}_{L_0,\ell_0}\right)_{ij}\prod_{(L,\ell)\in\cvec_i}\left({\bf N}_{L,\ell}\right)_{ii}\prod_{(L,\ell)\in\cvec_j}\left({\bf N}_{L,\ell}\right)_{jj}\,.\nonumber
\end{align}
The parameters  $L_0$ and $\ell_0$ and the parameter sets $\cvec_i$ and $\cvec_j$ characterize a process motif for steady-state correlation. 
A process motif for steady-state correlation consists of a process motif for steady-state covariance with focal nodes $i$ and $j$, a number $|\cvec_i|\geq0$ of process motifs for steady-state variance with positive length and focal node $i$, and a number $|\cvec_j|\geq 0$ of process motifs for steady-state variance with positive length and focal node $j$. We show a diagram of a process motif that contributes to correlation in \cref{fig:covcorr}\panel{C}. Both $|\cvec_i|$ and $|\cvec_j|$ can be equal to $0$, so process motifs for covariance are also process motifs for correlation.

\begin{figure*}[t]
\centering
\includegraphics[trim={1.575cm 17.7cm 3.3cm 3.8cm},clip,width=1\textwidth]{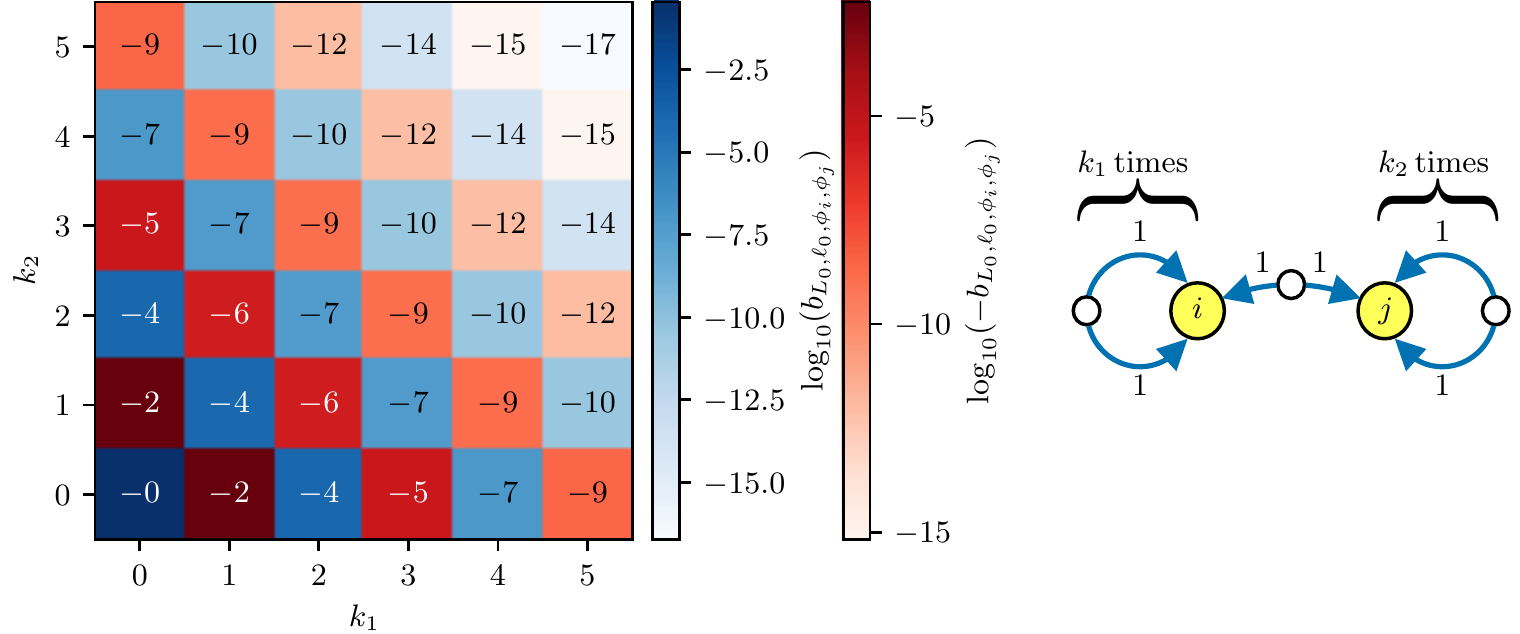} 
\caption{Contributions $b_{L_0,\ell_0,\cvec_i,\cvec_j}$ of process motifs to steady-state correlation with parameters $(L_0,\ell_0, \cvec_i,\cvec_j)$ for $\theta=1$, $\varsigma=1$, and $\epsilon=0.49$. We show the contributions for correlation process motifs that consist of a covariance process motif between node $i$ and node $j$, a number $n_i$ of variance process motifs at node $i$, and a number $n_j$ of variance process motifs at node $j$. All of the process motifs for variance and covariance have parameter values of $L=2$ and $\ell=1$.}
\label{fig:ccorr}
\end{figure*}

A contribution $b_{L_0,\ell_0,\cvec_i,\cvec_j}$ has a real non-zero value. All process motifs for correlation affect correlations, but not all process motifs for correlation contribute positively to correlations. The magnitude of $b_{L_0,\ell_0,\cvec_i,\cvec_j}$ is proportional to the contributions $b_{L,\ell}$ of the included process motifs for variance and covariance. One can construct process motifs for correlation with large contributions $b_{L_0,\ell_0,\cvec_i,\cvec_j}$ from process motifs for variance and covariance with large contributions $b_{L,\ell}$.

To illustrate the effect of the number of included variance process motifs on the contribution of a process motif to steady-state correlation, we show contributions of different process motifs to correlation in \cref{fig:ccorr}. The sign of $b_{L_0,\ell_0,\cvec_i,\cvec_j}$ is positive when the overall number of included process motifs for variance is even, and it is negative otherwise. The magnitude of $b_{L_0,\ell_0,\cvec_i,\cvec_j}$ decreases as one adds more process motifs for variance at either of the two focal nodes ($i$ and $j$). For a given process-motif length, the process motifs that contribute most to correlation are process motifs that do not include any process motifs for variance and are thus identical to process motifs for covariance. 
The process motifs with the largest negative contribution to correlation consist of a process motif for covariance and one process motif for variance at one of the focal nodes. 

These results match the intuition that (1) the correlation between two nodes $i$ and $j$ should increase with increasing covariance between $i$ and $j$ and 
(2) the correlation between them should decrease with increasing variance at either $i$ or $j$. The checkerboard structure of positive and negative contributions in \cref{fig:ccorr} arises because of the Taylor-series expansion for $1/\sqrt{x}$ in \cref{eq:expansion}. The summands in \cref{eq:expansion} have alternating signs, and under the assumption of short-range signal decay, the absolute value of the summands is strictly decreasing with $k$. In our derivation of $b_{L_0,\ell_0,\cvec_i,\cvec_j}$, we applied \cref{eq:expansion} twice (once for $1/\sqrt{\covel_{ii}}$ and once for $1/\sqrt{\covel_{jj}}$). This approach led to a sum over two indices, $k_1$ and $k_2$, where $k_1$ corresponds to the number of variance process motifs at node $i$ and $k_2$ corresponds to the number of variance process motifs at node $j$. Consequently, the sign of each summand in \eq{bcorr} and thus the sign of a process-motif contribution depends on the sum $k_1+k_2$ (which is equal to $|\phi_i|+|\phi_j|$ in \eq{bcorr}), and it alternates as one increases $k_1$ while keeping $k_2$ fixed (and vice versa). The absolute value of the contributions decreases with $k_1+k_2$, because short-range signal decay guarantees that the absolute value of the summands is strictly decreasing with increasing $k_1$ for any fixed $k_2$ and with increasing $k_2$ for any fixed $k_1$.


\subsection{Contributions of structure motifs to covariance and correlation at steady state}\label{sec:cov_contributions}

We now link the process motifs from \cref{sec:process_motifs} to network structure. In \cref{sec:contributions}, we defined the \textit{total contribution} $c_s$ of a structure motif $s$ as the sum of all contributions $b_{p}$ of all process motifs $p$ that occur on $s$ (see \eq{total_c}) and the \textit{specific contribution} $\hat c_s$ as the sum of contributions $b_{p}$ of process motifs $p$ that occur on $s$ but not on any subgraph of $s$ (see \eq{recurs}). For graphlets of up to six edges, we compute the total contributions and the specific contributions to covariance and correlation in the mOUP at steady state. We first demonstrate that one can explain most of the variation in the total contributions of structure motifs using the total contributions of their subgraphs. We then use the specific contributions of structure motifs to infer mechanisms by which network structure can contribute to covariance and correlation in the mOUP, and we compare the efficiency of these mechanisms.


\subsubsection{Total contributions of structure motifs to steady-state covariance}

\begin{figure*}[t]
\centering
\includegraphics[trim={2.2cm 1.5cm 2.525cm 0.9cm},clip,width=1\textwidth]{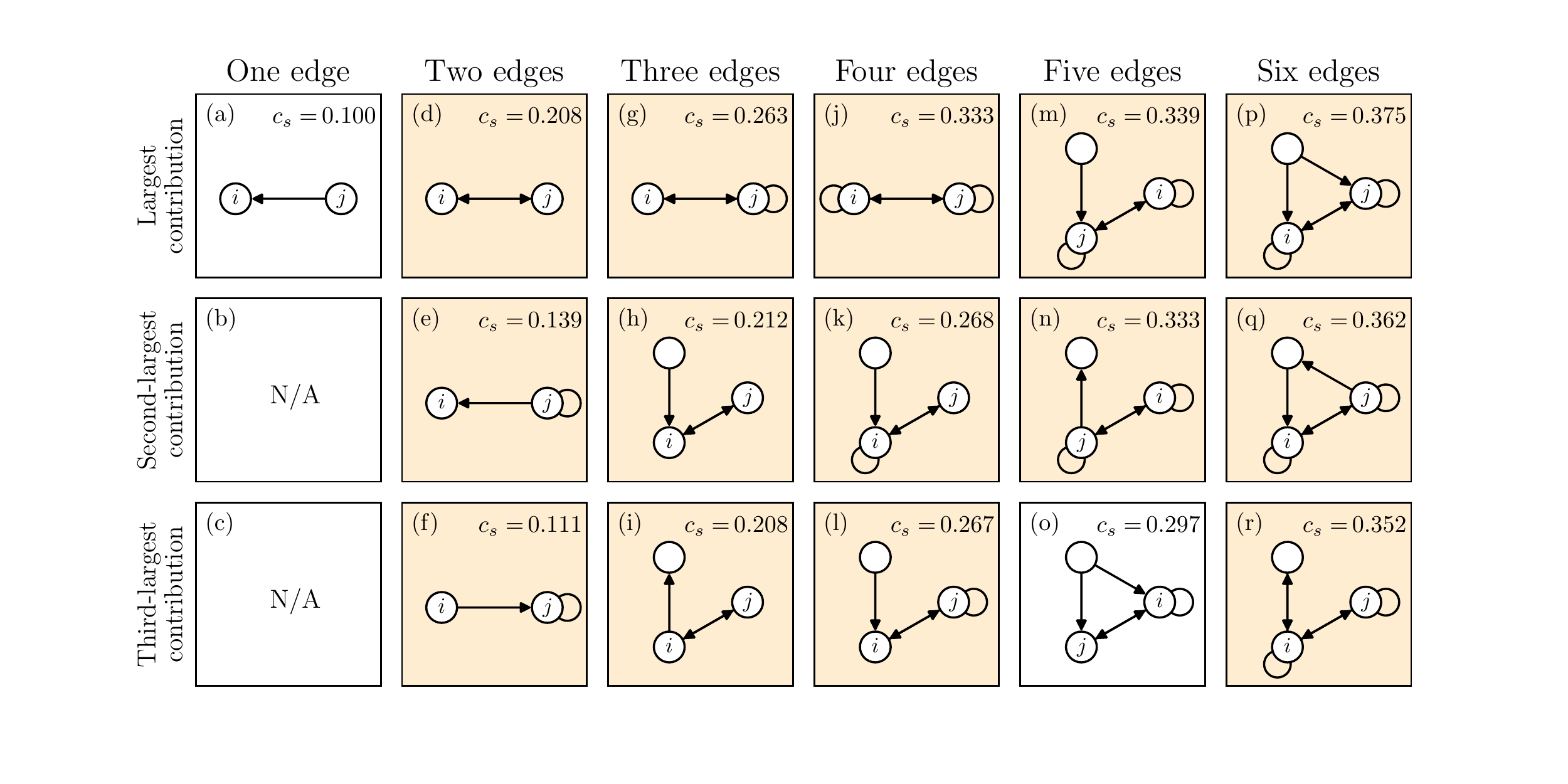} 
\caption{Structure motifs that have the largest total contribution $c_s$ to the steady-state covariance between nodes $i$ and $j$ in the mOUP (see \eq{OU}) with $\theta=1$, $\varsigma=1$, and $\epsilon=0.49$. To ensure that all adjacency matrices satisfy $\|{\bf A}\|\leq1$, we normalize each adjacency matrix by multiplying it by $1/\sqrt{6}$. We round the displayed values of $c_s$ to the third decimal place. Each panel with a peach background shows an $m$-edge structure motif that is a supergraph of the $(m-1)$-edge structure motif with the largest total contribution.}
\label{fig:cov_total}
\end{figure*}

In \cref{fig:cov_total}, we show the $m$-edge structure motifs with the three largest total contributions to covariance for $m\in\{1,2,\dots,6\}$. (Readers can explore the total and specific contributions of additional structure motifs using the Jupyter notebook in the Supplementary Materials \cite{SInote}.) There are many aspects of the structure motifs for covariance and their total contributions that one can explore. We focus on two results: (1) one can explain almost the entire variation in $c_s$ for structure motifs with $m$ edges using the mean total contribution $\langle c_{s'}\rangle_{m-1}$ of subgraphs with $m-1$ edges; and (2) process motifs are helpful for explaining salient properties of the structure motifs in \cref{fig:cov_total}.

\paragraph{Total contributions of subgraphs explain a large portion of the variation in the total contributions of structure motifs}
From \cref{fig:cov_total}, we see that, at least up to $m=6$, the three structure motifs with the largest total contributions are almost always supergraphs of the $(m-1)$-edge structure motif with the largest total contribution. This observation suggests that total contributions of subgraphs of a structure motif $s$ have a strong influence on the total contribution of $s$. To investigate the relationship between the total contributions of structure motifs and the total contributions of their subgraphs, we compute the Pearson correlation coefficient between $c_s$ of structure motifs with $m$ edges and the mean total contribution of their subgraphs with $m-1$ edges. We show the correlation coefficients in Table \ref{tab:mean_c}. We observe that one can explain almost all of the variation in $c_s$ using $\langle c_{s'}\rangle_{m-1}$. All of the correlation coefficients in Table \ref{tab:mean_c} are very large, and they increase with the number $m$ of edges and decrease with the mOUP coupling parameter $\epsilon$. 

To explain the large positive correlations between $c_s$ and $\langle c_{s'}\rangle_{m-1}$, we recall our discussion of the relationship between the total contributions of a structure motif and its subgraphs in \cref{sec:contributions}. Many process motifs that contribute to the total contribution $c_s$ of a structure motif $s$ do not use all edges in $s$ and are thus process motifs that also occur on subgraphs of $s$. Only process motifs that use every edge in $s$ cannot occur on any of the subgraphs of $s$. For a structure motif with $m$ edges, such a process motif has to have a length of $L\geq m$. We observe slightly decreasing correlation coefficients with increasing $\epsilon$ because contributions of long process motifs (e.g., process motifs with $L\geq m$) increase more than short process motifs with increasing $\epsilon$.

\begin{table}[h!]
\begin{center}
\begin{tabular}{c l l l l}
\toprule
 \multirow{2}{*}{\,\,\,$m$\,\,\,}   & \multicolumn{2}{c}{Covariance\,\,\,\,\,\,\,} & \multicolumn{2}{c}{Correlation}
\\
& $\epsilon=0.1$ \hspace{0.4cm}
& $\epsilon=0.49$ \hspace{0.4cm}
& $\epsilon=0.1$ \hspace{0.4cm}
& $\epsilon=0.49$ \hspace{0.4cm}\\
\midrule
2 
& 0.9985
& 0.9464
& 0.9993
& 0.9806
\\3 
& 0.9998
& 0.9932
& 0.9999
& 0.9966
\\4 
& 0.9999
& 0.9981
& $>0.9999$
& 0.9990
\\ 5 
& $>0.9999$
& 0.9993
& $>0.9999$
& 0.9996
\\6 
& $>0.9999$
& 0.9996
& $>0.9999$
& 0.9998
\\
\bottomrule
\end{tabular}
\end{center}
\caption{Pearson correlation coefficients between the total contributions $c_s$ of $m$-edge structure motifs to steady-state covariance in the mOUP and the mean total contributions $\langle c_{s'}\rangle_{m-1}$ of subgraphs with $m-1$ edges for different values of the mOUP coupling parameter $\epsilon$. For all of the coefficients that we show, the p-values are less than $10^{-17}$. 
}
\label{tab:mean_c}
\end{table}


\paragraph{Process motifs explain the properties of structure motifs with large total contributions}

\begin{figure*}[t]
\centering
\includegraphics[trim={2.52cm 9.35cm 2.0cm 1.75cm},clip,width=1\textwidth]{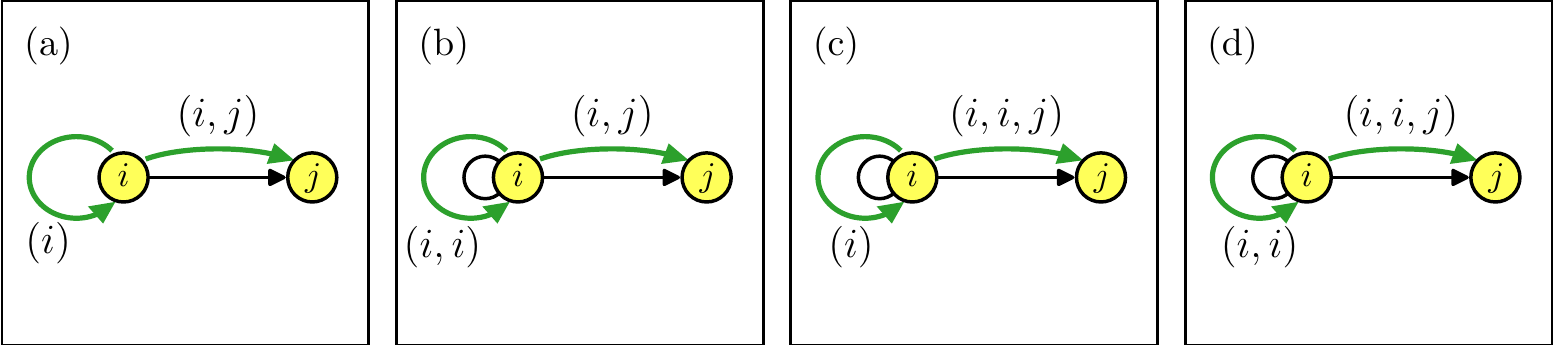} 
\caption{The effect of including a self-edge in a structure motif. In \panel{A}, we show a structure motif $s$ with one edge. The curved green edges indicate the only occurrence of the only process motif for covariance that occurs on $s$. In \panel{B}--\panel{D}, we show a supergraph of $s$ with a self-edge at node $i$. In each panel, the labeled green curved arrows indicate an occurrence of one of the many process motifs for covariance that occur on this graph, but not on $s$. All of the process motifs that occur include one walk from $i$ to $i$ and one walk from $i$ to $j$. The process motifs differ in the number of times that their walks visit $i$ and/or $j$.
}
\label{fig:self_edge_addition}
\end{figure*}

In thirteen of the sixteen structure motifs in \cref{fig:cov_total}, the focal nodes are connected bidirectionally. Twelve of the structure motifs in \cref{fig:cov_total} include self-edges. We first explain the high frequency of structure motifs with self-edges at focal nodes. Consider a structure motif $s$ that does not have a self-edge at either focal node. The inclusion of a self-edge at a focal node in $s$ yields a structure motif $s''$ that is a supergraph of $s$. Because $s''$ is a supergraph of $s$, every process motif that occurs on $s$ can also occur on $s''$. Additionally, for every length-$L$ process motif that occurs on $s$, there exist at least two process motifs with length $L+k$ on $s''$ for each $k\in\{1,2,3,\dots\}$. To illustrate this effect of including a self-edge in a structure motif, we show a simple example of a structure motif $s$ and a corresponding supergraph $s''$ with a self-edge in \cref{fig:self_edge_addition}. In \panel{A}, we show the only occurrence of the only covariance process motif that occurs on $s$. The process motif has a length of $1$. In \panel{B} and \panel{C}, we show the occurrences of the two process motifs with length 2 that occur on $s''$. In \panel{D}, we show an occurrence of one of the length-3 process motifs that occur on $s''$.

\begin{figure*}[!b]
\centering
\includegraphics[trim={2.52cm 9.35cm 2.0cm 1.75cm},clip,width=1\textwidth]{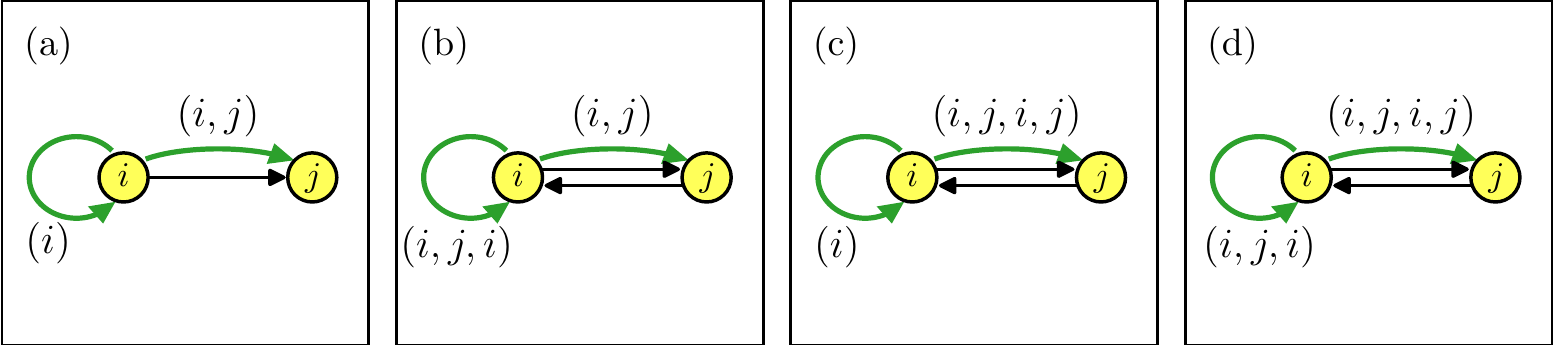} 
\caption{The effect of including a bidirectional edge in a structure motif. In \panel{A}, we show a structure motif $s$ with one edge. The curved green edges indicate the only occurrence of the only process motif for covariance that occurs on $s$. In \panel{B}--\panel{D}, we show a supergraph of $s$ with an edge $(j,i)$. In each panel, the labeled green curved arrows indicate an occurrence of one of the many process motifs for covariance that occur on this graph, but not on $s$.
The example process motifs in this figure include one walk from $i$ to $i$ and one walk from $i$ to $j$. The process motifs differ in the number of times that their walks visit $i$ and/or $j$.
}
\label{fig:edge_addition}
\end{figure*}

To explain the high frequency of structure motifs with bidirectionally connected focal nodes, we consider a structure motif $s$ with a unidirectional edge $(i,j)$ between focal nodes. The inclusion of an edge $(j,i)$ in $s$ yields a structure motif $s''$ that is a supergraph of $s$ and includes bidirectional coupling between its focal nodes. Because $s''$ is a supergraph of $s$, every process motif that occurs on $s$ can also occur on $s''$. For every process motif with length $L$ on $s$, there also exist at least two process motifs with length $L+2k$ for $k=1,2,3,\dots$. We illustrate the effect of including a bidirectional edge in a structure motif in \cref{fig:edge_addition}. In \panel{A}, we again show the structure motif from the example in \cref{fig:self_edge_addition} and the occurrence of the length-$1$ process motif that is the only covariance process motif that occurs on $s$. In \panel{B} and \panel{C}, we show the occurrences of the two process motifs with length $3$ that occur on $s''$. In \panel{D}, we show an occurrence of one of the length-$5$ process motifs that occur on $s''$.

The high frequencies of self-edges and edges between focal nodes in structure motifs that contribute the most to steady-state covariance and correlation suggest that signal transmission via short paths between focal nodes and signal amplification via short cycles are important for mechanisms by which network structure can contribute to covariances in the mOUP.


\subsubsection{Specific contributions of structure motifs to covariance}\label{sec:covspecific}

In \cref{sec:part2}, we proposed to separate the total contribution of a structure motif $s$ into a large portion that one can attribute to subgraphs of $s$ and a small portion that one cannot attribute to subgraphs of $s$. The small portion $\specific_s$ is the specific contribution of $s$. The specific contribution of the structure motif with one edge indicates the contribution to covariance of a single edge. The specific contributions of structure motifs with two edges indicate the contribution to covariance of a pair of edges minus the specific contributions of each of the two edges alone. Whenever the specific contribution of a structure motif is positive, the structure motif indicates a mechanism or a combination of mechanisms by which network structure can enhance covariance.


\paragraph{Structure motifs with $\specific_s>0$ indicate mechanisms for structure-based enhancement of steady-state covariance}

\begin{figure*}[t] 
\centering
\includegraphics[trim={3.14cm 0.9cm 2.53cm 0.95cm},clip,width=1\textwidth]{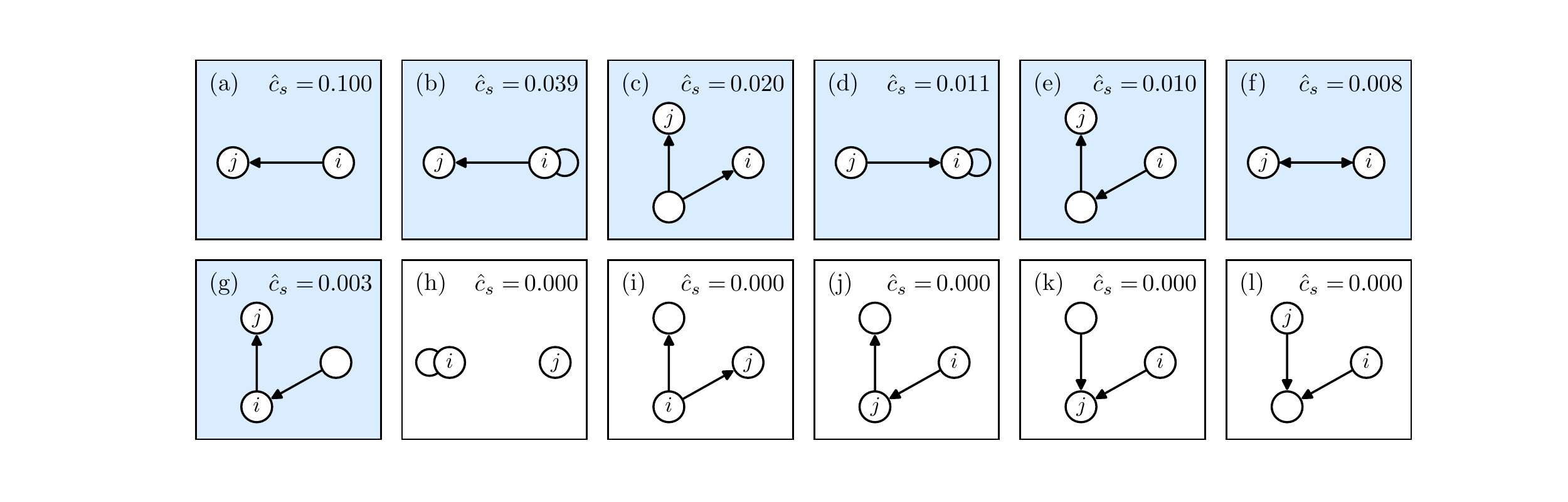} 
\caption{Structure motifs with one or two edges and their specific contributions $\specific_s$ to covariance in the mOUP (see \eq{OU}) with $\theta=1$, $\varsigma=1$, and $\epsilon=0.49$. To ensure that all adjacency matrices satisfy $\|{\bf A}\|\leq1$, we normalize each adjacency matrix by multiplying it by $1/\sqrt{6}$. We round the displayed values of $\specific_s$ to the third decimal place. Panels \panel{A}--\panel{G} have blue backgrounds and show structure motifs with a positive $\specific_s$.}
\label{fig:cov12}
\end{figure*}

In \cref{fig:cov12}, we show structure motifs with one or two edges and their specific contributions to steady-state covariance. Panels \panel{A}--\panel{G} have blue backgrounds and show structure motifs with a positive specific contribution. These structure motifs indicate mechanisms for enhancing steady-state covariance in the mOUP. In panel \panel{H}, we include a graphlet that is not a structure motif because it has two components. We include it because it is helpful for discussing the mechanisms by which network structure can contribute to covariance in the mOUP.  The positive specific contributions for structure motifs in panels \panel{A} and \panel{E} indicate that signal transmission via short paths from one focal node to the other can increase covariance. The positive specific contributions for structure motifs in panels \panel{B} and \panel{D} indicate that signal amplification via a length-$1$ cycle can increase covariance when combined with a path for signal transmission between focal nodes. The specific contribution of the small graph in panel \panel{H} is $0$, which indicates that signal amplification at a focal node does not increase covariance without any connection between focal nodes. In panel \panel{F}, the bidirectional edge between focal nodes enables signal transmission from either focal node to the other. It also creates a 2-cycle at each focal node. The positive specific contributions for structure motifs in panels \panel{C} and \panel{G} indicate that a signal transmission from a non-focal node can contribute to covariance. Comparing panels \panel{C} and \panel{G} to panel \panel{K}, we see that a positive $\specific_s$ requires that there exist paths from the non-focal node to both focal nodes. The $0$ contributions of the structure motifs in panels \panel{I}, \panel{J}, and \panel{L} indicate that paths from focal nodes to other nodes are not relevant for the covariance between focal nodes.

From these observations, we conclude that two mechanisms for increasing steady-state covariance in the mOUP are (1) signal transmission via paths from one focal node to another and (2) signal transmission via paths from non-focal nodes to each focal node. Other mechanisms for increasing covariance in the mOUP are combinations of signal transmission via paths between focal nodes and signal transmission from non-focal nodes. Such mechanisms can also be combinations of either or both mechanisms with signal amplification via short cycles at focal nodes or other nodes. 


\paragraph{Specific contributions indicate the efficiency of mechanisms}

Thus far, we have used specific contributions to distinguish structure motifs that contribute to steady-state covariance (i.e., structure motifs with $\specific_s>0$) from structure motifs that do not (i.e., structure motifs with $\specific_s=0$). We can use the value of specific contributions to define a measure of mechanism efficiency. For a structure motif with $m$ edges and specific contribution $\specific_s$, we define the efficiency
\begin{align}
	\effi:=\specific_s/m\,.\nonumber
\end{align}

From \cref{fig:cov12}, we see that specific contributions and thus $\effi$ tend to decrease with the number of edges in a structure motif. The mechanisms with large efficiency tend to be associated with small structure motifs. The structure motif with the largest specific contribution to covariance (see \cref{fig:cov12}\panel{A}) indicates direct signal transmission (i.e., signal transmission via a length-$1$ path) as a mechanism for increasing covariance. The associated efficiency is $\effi \approx 0.1$. All other mechanisms have much smaller efficiencies than direct signal transmission. For example, signal transmission via a length-$2$ path (see \cref{fig:cov12}\panel{E}) has an efficiency of $\effi \approx 0.005$, and ones through longer paths are even smaller.

When the focal nodes are connected by a single directed path, one can think of the focal node with positive out-degree as the ``sender'' node and the node with positive in-degree as the ``receiver'' node. The second-most efficient mechanism is a combination of direct signal transmission and signal amplification via a length-$1$ cycle at the sender node (see \cref{fig:cov12}\panel{B}). This mechanism has an efficiency of $\effi\approx0.02$. The efficiency of direct signal transmission with signal amplification via a length-$1$ cycle at the receiver node (see \cref{fig:cov12}\panel{D}) has an efficiency of $\effi\approx0.005$, which is almost four times smaller than the efficiency of the mechanism in \cref{fig:cov12}\panel{B}.
Transmission of signals from a third node to both focal nodes via length-$1$ paths (see \cref{fig:cov12}\panel{C}) has an efficiency of $\effi \approx 0.01$. 


\paragraph{Matching motifs give a heuristic way to explain specific contributions}

For the mechanisms that are associated with $1$-edge and $2$-edge structure motifs, one can explain the ranking of specific contributions using matching process motifs. If a process motif $p$ contributes to the specific contribution $\specific_s$ of $s$, it uses each edge in $s$ at least once; otherwise, it would contribute to the specific contribution of a proper subgraph of $s$ and not to the specific contribution of $s$. The contributions of process motifs tend to decrease with their length. Therefore, the largest contributions of process motifs to $\specific_s$ of $s$ come from matching process motifs of $s$. One can use the sum 
\begin{align}
	\gamma_s:=\sum_{p\in P^*_S} b_p \nonumber
\end{align}
 of contributions of matching process motifs as a heuristic for estimating $\specific_s$. We show the Pearson correlation coefficients for $\specific_s$ and $\gamma_s$ for different structure-motif lengths in Table \ref{tab:pearson}. As a comparison, we also show a second heuristic $\gamma_s'=\max_{p\in P^*_S} b_p$ that only uses the contribution of the matching process motifs that contributes the most to covariance. We observe that there is a large positive correlation between $\specific_s$ and $\gamma_s$ for all considered structure-motif lengths. The heuristic $\gamma_s$ is correlated most strongly with $\specific_s$ when both $m$ and $\epsilon$ are small. The heuristic $\gamma_s'$ also has a large positive correlation with $\specific_s$ for $m=2$. However, as we consider structure motifs with progressively more edges, the Pearson correlation coefficient between $\specific_s$ and $\gamma_s'$ decreases much faster than the Pearson correlation coefficient between $\specific_s$ and $\gamma_s$. This difference between the two heuristics demonstrates that it is important to consider all matching process motifs instead of just one matching process motif for structure motifs with more than two edges.

\begin{table}[h!]
\begin{center}
\begin{tabular}{c l l l l} 
\toprule
 \multirow{2}{*}{\,\,\,$m$\,\,\,}   & \multicolumn{2}{c}{Covariance\,\,\,\,\,\,\,} & \multicolumn{2}{c}{Correlation}
\\
& $\epsilon=0.1$ \hspace{0.4cm}
&  $\epsilon=0.49$ \hspace{0.4cm}
& $\epsilon=0.1$ \hspace{0.4cm}
& $\epsilon=0.49$ \hspace{0.4cm}\\
\midrule
2 
& 0.998  
& 0.966 
& 0.962
& 0.900
\\3 
& 0.996  
& 0.903 
& 0.958
& 0.855
\\4 
& 0.994  
& 0.814 
& 0.913
& 0.723
\\5 
& 0.996
& 0.879 
& 0.854
& 0.718
\\6 
& 0.993 
& 0.820 
& 0.811
& 0.606
\\
\bottomrule
\end{tabular}
\end{center}
\caption{Pearson correlation coefficients for the heuristics $\gamma_s$ and $\gamma_s'$ with specific contribution $\specific_s$ of $m$-edge structure motifs to steady-state covariances of the mOUP with a coupling parameter of $\epsilon$. For all of these Pearson correlation coefficients, the p-values are less than $0.00017$.}
\label{tab:pearson}
\end{table}


\subsubsection{Specific contributions of structure motifs to steady-state correlation}

\begin{figure*}[t]
\centering
\includegraphics[trim={2.2cm 1.5cm 2.525cm 0.9cm},clip,width=1\textwidth]{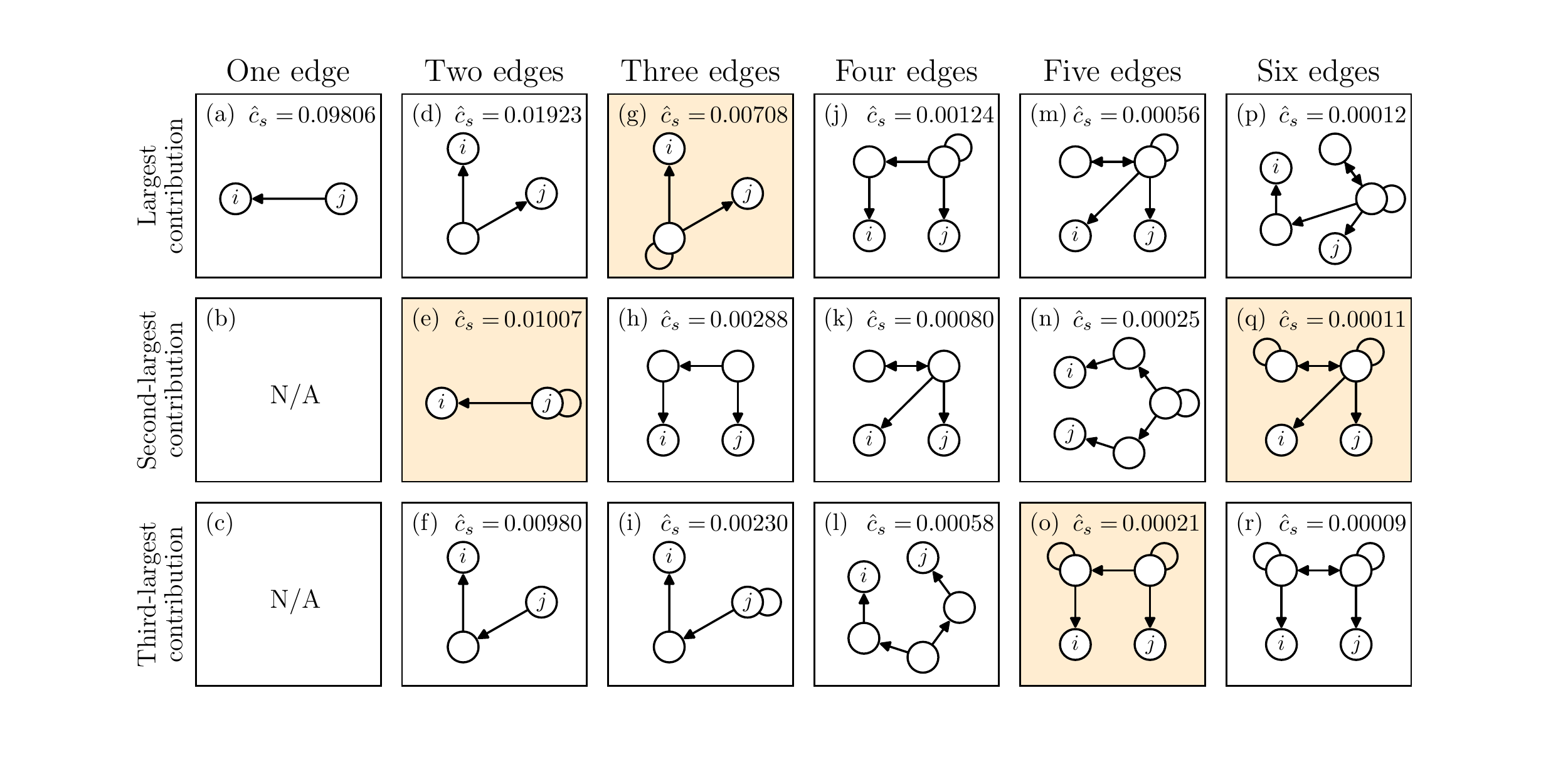} 
\caption{Structure motifs that have the largest specific contributions $\specific_s$ to the correlation between nodes $i$ and $j$ in the mOUP (see \eq{OU}) with $\theta=1$, $\varsigma=1$, and $\epsilon=0.49$. To ensure that all adjacency matrices satisfy $\|{\bf A}\|\leq1$, we normalize each adjacency matrix by multiplying it by $1/\sqrt{6}$. We round the displayed values of $\specific_s$ to the fifth decimal place. Each panel with a peach background shows an $m$-edge structure motif that is a supergraph of the $(m-1)$-edge structure motif with the largest specific contribution $\specific_s$.}
\label{fig:cor_specific}
\end{figure*}

We demonstrated in \cref{sec:cov_contributions} that the specific contributions of structure motifs convey covariance-enhancing mechanisms more 
clearly than total contributions. In this section, we focus on the specific contributions of structure motifs to correlations in the mOUP. In \cref{fig:cor_specific}, we show the $m$-edge structure motifs with the three largest specific contributions to steady-state correlation for $m\in\{1,2,\dots,6\}$. Readers can explore the total and specific contributions of further structure motifs using the Jupyter notebook in the Supplementary Materials \cite{SInote}.


\paragraph{Network structure can increase or decrease steady-state correlations}

\begin{figure*}[t] 
\centering
\includegraphics[trim={3.14cm 0.9cm 2.53cm 0.95cm},clip,width=1\textwidth]{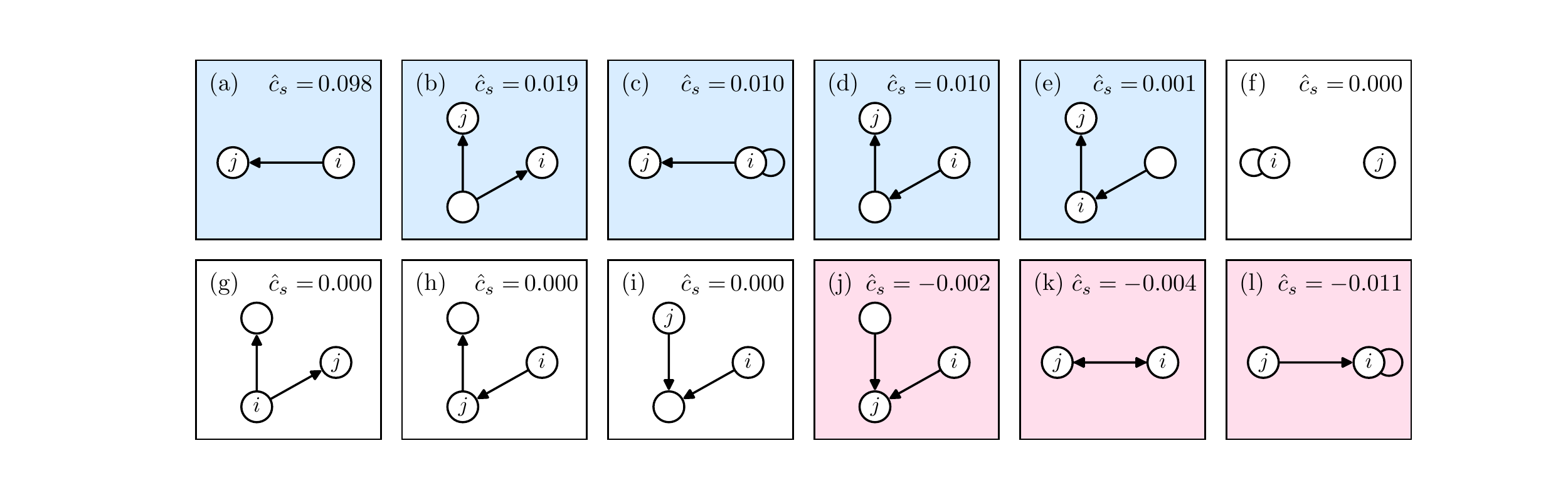} 
\caption{Structure motifs with one or two edges and their specific contributions $\specific_s$ to the steady-state correlation in the mOUP (see \eq{OU}) with $\theta=1$, $\varsigma=1$, and $\epsilon=0.49$. To ensure that all adjacency matrices satisfy $\|{\bf A}\|\leq1$, we normalize each adjacency matrix by multiplying it by $1/\sqrt{6}$. We round the displayed values of $\specific_s$ to the third decimal place. Panels \panel{A}--\panel{E} show structure motifs with a positive $\specific_s$ and have blue backgrounds. Panels \panel{J}--\panel{L} show structure motifs with a negative $\specific_s$ and have pink backgrounds.}
\label{fig:cor12}
\end{figure*}

In \cref{fig:cor12}, we show structure motifs with one or two edges and their specific contributions to the steady-state correlation in the mOUP. 
Negative specific contributions to correlation in the mOUP indicate that there are mechanisms by which network structure can decrease the correlation between a pair of nodes in the mOUP. The structure motifs with negative $\specific_s$ for correlation include structure motifs that have a $0$ specific contribution to covariance. An example is the structure motif in \cref{fig:cor12}\panel{J}. Its specific contribution to covariance is $\specific_s=0$, from which we concluded in \cref{sec:covspecific} that signal transmission from a non-focal node to only one focal node does not increase covariance in the mOUP. The same structure motif has a negative specific contribution to correlation in the mOUP. From this negative specific contribution, we conclude that signal transmission from a non-focal node to a single focal node can decrease correlation in the mOUP. 

The decrease in correlation via this mechanism arises because of the normalizing factor $1/\sqrt{\covel_{ii}\covel_{jj}}$ in the definition of the correlation coefficient $\correl_{ij}$ (see \cref{eq:rij}). Signal transmission from non-focal nodes to a single focal node increases the variance in that focal node without increasing the covariance between the pair of focal nodes. The correlation between a pair of nodes is inversely proportional to the variance at each node. Consequently, an increase of variance in one focal node without a compensating increase of the covariance between the pair of focal nodes leads to a decrease of the correlation between them. Intuitively, the states of two nodes $i$ and $j$ cannot be perfectly correlated if the node $i$ also receives and responds to signals from other nodes that are not connected to $j$. The more such signals that node $i$ receives, the more its correlation with $j$ decreases.

The structure motifs with negative $\specific_s$ for correlation also include structure motifs that have a positive $\specific_s$ for covariance. An example is the structure motif in \cref{fig:cor12}\panel{L}. In \cref{sec:covspecific}, we concluded that direct signal transmission with signal amplification at the receiver node is a mechanism by which network structure can increase covariance in the mOUP. From the structure motif's negative specific contribution to correlation, we conclude that (by the same mechanism) network structure can decrease correlation in the mOUP.

To give an intuitive explanation for the qualitative and quantitative differences in efficiency of direct signal transmission with amplification at the sender node or the receiver node, we contrast the effect of an amplifier at a sender and at a receiver in a system with additive noise. On one hand, amplifying a signal at a sender node increases the amplitude of the signal, improves the signal-to-noise ratio at the receiver node, and thus leads to an increase of covariance and correlation between the sender node and the receiver node (see \cref{fig:cov12}\panel{B} and \cref{fig:cor12}\panel{C}). On the other hand, amplifying a signal at a receiver node increases the amplitude of the signal and the noise at the receiver. Therefore, a signal amplification at the receiver does not change the signal-to-noise ratio at the receiver node and hence it does not increase correlation between the sender node and the receiver node. It does lead to a small increase of covariance (see \cref{fig:cov12}\panel{D}) and a small decrease of correlation (see \cref{fig:cor12}\panel{L}) between the two nodes through increases in the amplitudes of the signal and the noise (and hence of the variance) at the receiver node.


\paragraph{The influence of variance leads to different rankings of mechanisms for covariance and correlation}

Comparing the specific contributions to covariance and correlation for a given structure motif, we see that the contributions are almost identical for some structure motifs. For example, the structure motif in \cref{fig:cor12}\panel{A} has $\specific_s \approx 0.100$ for covariance and $\specific_s \approx 0.098$ for correlation. Another example is the structure motif in \cref{fig:cor12}\panel{B}. It has $\specific_s \approx 0.02$ for covariance and $\specific_s \approx 0.019$ for correlation. For other structure motifs, the specific contributions to correlation are much smaller than their specific contributions to covariance. For example, the structure motif in \cref{fig:cor12}\panel{C} has $\specific_s \approx 0.039$ for covariance and $\specific_s \approx 0.01$ for correlation. Some structure motifs have a non-negative specific contribution to covariance but a negative specific contribution to correlation. For example, we discussed earlier in this section that the structure motif in \cref{fig:cor12}\panel{K} has $0$ specific contribution to covariance but a negative specific contribution to correlation. We also discussed that the structure motif in \cref{fig:cor12}\panel{D} has $\specific_s \approx 0.011$ for covariance but $\specific_s \approx -0.011$ for correlation.

These differences between the specific contributions to covariance and correlation are related to the process motifs for variance in focal nodes. Structure motifs on which few process motifs for focal-node variance node occur tend to have very similar specific contributions to covariance and correlation. Examples of such structure motifs are the ones in \cref{fig:cor12}\panel{A} and \panel{B}. For other structure motifs, the specific contributions to covariance and correlation are very different, because many process motifs for variance at focal nodes occur on them. The structure motif in \cref{fig:cor12}\panel{C} is an example of such a structure motif. 

Because of these differences, ranking structure motifs by their specific contribution to covariance and ranking structure motifs by their specific contribution to correlation lead to different rankings. Consequently, rankings that are based on the efficiencies of the associated mechanisms are also different.


\paragraph{Increasing the in-degree of a receiver node reduces steady-state correlations in locally tree-like networks}

The structure motifs in \cref{fig:cor12}\panel{E}, \panel{G}, \panel{H}, and \panel{J} include a directed edge between focal nodes and an in-edge or out-edge at the sender node or the receiver node. We use these structure motifs and their specific contributions to correlation to study the effect of increasing the in-degree or out-degree of focal nodes on correlations in a locally tree-like network \cite{Melnik2011}. Because we are considering locally tree-like networks, we assume that neighbors of a node are not neighbors of each other. We also assume that we can neglect structure motifs with more than two edges because such structure motifs tend to have very small specific contributions (see \cref{fig:cor_specific}). Under these assumptions, we make the following observations:
\begin{enumerate}
\item an increase of the in-degree of a sender node leads to an increase of the count of the structure motif in \cref{fig:cor12}\panel{E} but of no other structure motifs;
\item an increase of the out-degree of a sender node leads to an increase of the count of the structure motif in \cref{fig:cor12}\panel{G} but of no others; 
\item an increase of the in-degree of a receiver node leads to an increase of the count of the structure motifs in \cref{fig:cor12}\panel{H} but of no others; and
\item an increase of the out-degree of a receiver node leads to an increase of the count of the structure motifs in \cref{fig:cor12}\panel{J} but of no others.
\end{enumerate}
One can infer the effect of increasing in-degree or increasing out-degree of the sender node or the receiver node from the specific contributions of these structure motifs. Increasing the out-degree of the sender node (see \cref{fig:cor12}\panel{G}) or the receiver node (see \cref{fig:cor12}\panel{F}) does not affect the correlation between the sender and the receiver. Increasing the in-degree of the sender node (see \cref{fig:cor12}\panel{E}) leads to an increase of the correlation (to $\eta\approx5\cdot10^{-4}$). Increasing the in-degree of the receiver node (see \cref{fig:cor12}\panel{J}) leads to a decrease of the correlation (to $\eta\approx-10^{-3}$).

When two focal nodes are connected bidirectionally, one cannot distinguish between a sender node and a receiver node. Increasing the in-degree of either focal node increases the counts of the structure motif in \cref{fig:cor12}\panel{E} and the structure motif in \cref{fig:cor12}\panel{J} by $1$ each. The net effect of increasing the in-degree of a focal node is given by the sum of the specific contributions of the structure motifs in \cref{fig:cor12}\panel{E} and \panel{J}. This sum is negative. Therefore, increasing the in-degree of a node in a locally tree-like network reduces the correlation between this node and nodes with which it is connected bidirectionally.


\section{Conclusions and discussion}\label{sec:conclusions}

Discovering connections between dynamics on networks and network structure is an ongoing endeavor in many disciplines. Many researchers find it helpful to decompose networks into structural building blocks, which are typically called ``motifs''. In the present paper, we demonstrated that combining such a decomposition of a network's structure into structure motifs with a decomposition of processes on a network into process motifs can yield both mechanistic and quantitative insights into connections between dynamics on networks and network structure. To construct a framework for the combined decomposition of processes on networks and network structure, we introduced process motifs as ``building blocks'' of processes and defined contributions of process motifs and total contributions and specific contributions of structure motifs to observables of dynamical systems on networks. 


\subsection{Mechanisms for enhancing steady-state covariance and steady-state correlation in the Ornstein--Uhlenbeck process}

To demonstrate our framework, we performed a combined decomposition into process and structure motifs for the multivariate Ornstein--Uhlenbeck process 
(mOUP) on a network. We identified the process motifs that contribute to variances, covariances, and correlations in the mOUP at steady state. We then used the contributions of the identified process motifs to variance, covariance, and correlation to explain the total contributions and specific contributions of structure motifs to covariance and correlation. The specific contributions of structure motifs signify several mechanisms by which network structure can enhance the covariance or the correlation between two focal nodes in the mOUP at steady state. Structure motifs contribute positively to covariance and correlation by enhancing signal transmission between focal nodes; signal transmission from non-focal nodes; or a combination of signal transmission between focal nodes, signal transmission from non-focal nodes, and signal amplification at focal nodes or non-focal nodes. The ranking of structure motifs and associated mechanisms by specific contributions is different for covariance and correlation, and it depends on the coupling parameter $\epsilon$ of the mOUP.

Some of our results on process motifs and structure motifs for covariance and correlation for the mOUP may match one's intuition for covariance and correlation. For example, the popular phrase ``correlation does not imply causation'' is consistent with our results that (1) process motifs for covariance and correlation between two nodes $i$ and $j$ do not necessarily include a walk from $i$ to $j$ or from $j$ to $i$ and (2) structure motifs for covariance and correlation do not necessarily include a path or trail from $i$ to $j$ or from $j$ to $i$. Our findings confirm known results about the mechanisms by which network structure can affect covariances and correlations between variables, and they also offer new quantitative insights into the efficiency of these mechanisms and the relationship between efficiency and the mOUP parameters. We anticipate that at least some of our findings hold also for other dynamical systems on networks. For example, the process motifs for covariance in the mOUP are equivalent to process motifs for the mean covariance in a system of coupled Hawkes processes \cite{Pernice2011} and the process motifs for coherence in a system of coupled integrate-and-fire neurons \cite{Hu2014}.


\subsection{Applicability to other dynamical systems}

In the present paper, we studied covariance and correlation in a simple stochastic dynamical system (specifically, the mOUP) at steady state. We chose this example for illustrative purposes and to demonstrate that our approach can confirm and extend intuition about the network mechanisms that contribute to system function. It is also possible to apply our framework to other system functions, other linear dynamical systems, and away from a steady state. For dynamical systems that are away from a steady state, we note that process-motif decompositions of system functions can depend on initial conditions. 

We considered structure motifs in directed networks with self-edges. For some systems in biology, chemistry, sociology, and other areas, it can be appropriate to consider undirected networks or networks without self-edges. One can apply our framework to such networks by focusing the motif comparison on structure motifs in undirected networks or networks without self-edges. Because of the flexibility of our approach, we anticipate that the study of process motifs can yield insights into many open problems in the study of dynamical systems on networks.


\subsection{When does the distinction between process motifs and structure motifs matter?\hspace{-0.158cm}}\label{sec:matter}

The distinction between process motifs and structure motifs matters for many dynamical systems, but it does not matter for all of them. Our core motivation for distinguishing between process motifs and structure motifs is that a walk on a network and a path in a network are two fundamentally different concepts. A walk can use an edge in a network several times, whereas a path or trail can include each edge only once. When one defines a process on a network such that it can use each node only once, the distinction between walks and paths becomes unnecessary because every path corresponds to a single walk. Examples of such processes include susceptible--infected (SI) models and susceptible--infected--recovered (SIR) models for the spread of an infectious disease \cite{Kiss2017,Porter2016}. Infected and recovered individuals in these models cannot become infected a second time, so a disease can spread along each edge at most once. One can construct other models that allow recurring infections (i.e., an individual can become infected multiple times). Examples of such models are susceptible--infected--susceptible (SIS) models and susceptible--infected--recovered--susceptible (SIRS) models. For such models, it is important to distinguish between process motifs and structure motifs. One can circumvent the need to make a distinction by introducing restrictive model assumptions that are popular in the modeling of infectious diseases \cite{Kiss2017}. For example, one can assume that 
\begin{enumerate}
\item a network is a directed acyclic graph (DAG) or 
\item a network is directed and locally tree-like and that infection rates are low.
\end{enumerate}
On a DAG, there are no process motifs that use an edge in a network more than once. Under assumption (1), the distinction between process motifs and structure motifs does not matter. In networks that are both directed and locally tree-like, there are no process motifs with length $L\leq3$ that use an edge in a network more than once. A low infection rate ensures that the contributions of long process motifs are very small. Under assumption (2), the distinction between process motifs and structure motifs has only a small effect on the specific contributions of structure motifs. We anticipate that distinguishing between process motifs and structure motifs can aid researchers in the study of diseases on networks using models that allow recurring infections.

When a network is a DAG, walks on it cannot use an edge more than once, so the distinction between process motifs and structure motifs is not relevant for any dynamical system on a DAG. There are numerous applications of dynamical systems on DAGs in machine learning and neuroscience \cite{Shinozaki2015, Tino2015}. They include feedforward artificial neural networks and models of natural neural networks in the visual cortex of several species \cite{Kampa2006, Vanrullen2003}. Many researchers in machine learning and neuroscience have highlighted the fundamental differences in the dynamics of non-recurrent neural networks (i.e., neural networks that are DAGs) and recurrent neural networks (i.e., those that are not DAGs) \cite{Manjunath2013, Tino2015}. We anticipate that our framework for decomposing processes on networks into process motifs can help explain some of these differences between non-recurrent and recurrent neural networks.

Dynamical systems on temporal networks are another example for which one can sometimes ignore the distinction between process motifs and structure motifs. 
One can define many temporal networks such that each edge is active only at a specified point in time or during a specified time interval \cite{Holme2019}. 
When each edge in a temporal network is active only at very few times points or only for time intervals that are short in comparison to the temporal scales of the processes on a network, few or no walks on the temporal network use an edge more than once. On such temporal networks, it is possible that each structure motif has only a few associated process motifs. (See \cite{Porter2016} for a discussion of the relative temporal scales of dynamics on networks and dynamics of networks.) The development of new notions of structure motifs in temporal networks is an active field of research, and researchers have made several proposals for notions of structure motifs in temporal networks \cite{Hulovatyy2015, Lehmann2019, Liu2020, Paranjape2017, Roldan2020, Tu2018}. The distinction between process motifs and structure motifs may be helpful for assessing these proposals and for the development of further notions of motifs on temporal networks.


\subsection{``Unbiased'' mechanistic insights from process motifs and structure motifs}

In this paper, we presented an approach for identifying graphlets that are relevant to a function of a system. Our approach offers several advantages over traditional approaches, in which researchers use overrepresentation of graphlets as a surrogate to conclude that graphlets are relevant to a system function. Those approaches depend strongly on the choice of a random-graph null model \cite{Artzy-Randrup2004, Robin2007, Schlauch2015}, and they do not identify mechanisms by which overrepresented graphlets affect a system function. Studies of dynamical systems on graphlets in isolation require researchers to choose a graphlet and a candidate mechanism \textit{a priori}. The reliance on these choices makes such studies prone to bias towards graphlets or mechanisms that a researcher has chosen to study. For example, many studies have reported the relevance of feedback loops and feedforward loops to various system functions \cite{Alon2007}. However, it is unclear if these two graphlets are generally more important for system functions than other graphlets or if researchers have associated them more frequently than other graphlets with system function because they have studied them more often. 

Our approach identifies all structure motifs with a positive (or a negative) contribution to a given function of a dynamical system. The approach is unbiased in the sense that its results do not depend on an \textit{a priori} choice of a graphlet or a mechanism. Our results for steady-state covariance and steady-state correlation in the mOUP demonstrate that there can be many structure motifs that affect a system function. Had we considered only a single graphlet in our study, it is likely that we would have concluded that that graphlet affects steady-state covariance and steady-state correlation in the mOUP and would then have inferred that that graphlet is important for these system functions. Our systematic study of all graphlets with up to six edges enabled us to rank structure motifs based on their contributions and also made it possible to distinguish between structure motifs that strongly affect steady-state covariance and steady-state correlation in the mOUP and structure motifs that have smaller (or even negligible) contributions to these system functions.

We also demonstrated how to perform a combined decomposition of dynamics on a network and network structure into process motifs and structure motifs. One can use such a decomposition to identify the structure motifs that contribute the most to a given system function and to explain how these structure motifs contribute to the system function. We demonstrated that it can be useful to consider dynamics on a network (instead of just a network's structure) as a composite object that one can decompose into many small parts. Our proposed framework thereby provides an opportunity to develop insights into mechanisms by which dynamics and network structure affect system functions.


\appendix

\section{Derivation of a non-recursive formula for the specific contributions of structure motifs}\label{sec:app:direct}
Consider a structure motif $s$ with $m$ edges and specific contribution $\specific_s$. Successive recursions of \eq{recurs} lead to an expression that depends only on the total contributions $c_{s'}$ of subgraphs $s'$ of $s$. Subgraphs with the same number $m'<m$ of edges contribute to $\specific_s$ in the same way. One can thus write
\begin{align}
	\specific_s = \alpha_1\langle c_{s'}\rangle_{1}+\alpha_2\langle c_{s'}\rangle_{2} + \alpha_3\langle c_{s'}\rangle_{3} + \cdots + \alpha_{m-1}\langle c_{s'}\rangle_{m-1}\,+\alpha_m\langle c_{s'}\rangle_{m}\,,\label{eq:alphasum}
\end{align}
where $\alpha_{\indA}$ are integer-valued coefficients with indices $\indA\in\{1,\dots,m\}$ and $\langle c_{s'}\rangle_{\indA}$ is the mean total contribution of the subgraphs of $s$ with $\indA$ edges. The structure motif $s$ has exactly one subgraph (specifically, the graph itself) with $m$ edges, so $\langle c_{s'}\rangle_{m}=c_s$. From \eq{recurs}, we see that a structure motif $s$ with one edge (i.e., the ``0-th'' recursion of \eq{recurs}) has $\alpha_m=1$. Further recursions of \eq{recurs} do not change $\alpha_m$, because subgraphs with $m$ edges are not proper subgraphs of $s$ and thus do not appear in the sum over proper subgraphs $s'\subset s$ in \eq{recurs}. The first recursion of \eq{recurs} yields
\begin{align}
	\specific_s = c_s-\sum_{s'_1\subset s} \left(c_{s'_1}-\sum_{s'_2\subset s'_1}\specific_{s'_2}\right)\label{eq:recurs2}\,.
\end{align}
From \eq{recurs2}, we see that $\alpha_{m-1}$ is equal to the negative of the number of subgraphs of $s$ with $m-1$ edges in the first recursion of \eq{recurs}. Further recursions of \eq{recurs} do not change $\alpha_{m-1}$, because subgraphs with $m-1$ edges cannot be proper subgraphs of proper subgraphs $s'_1$ of $s$ and thus do not appear in the sum over proper subgraphs $s'_2\subset s'_1$. It thus follows that
\begin{align}
	\alpha_{m-1}=-\binom{m}{m-1}\,.\nonumber
\end{align}
Subgraphs with $m-2$ edges are proper subgraphs of $s$. We thus obtain $\alpha_{m-2}=-\binom{m}{m-2}$ in the first recursion of \eq{recurs}. Because subgraphs with $m-2$ edges are also proper subgraphs of proper subgraphs $s'_1$ of $s$, the second recursion of \eq{recurs} leads to an additional term $\binom{m}{m-1}\binom{m-1}{m-2}$ in $\alpha_{m-2}$. Further recursions of \eq{recurs} do not change $\alpha_{m-2}$. It thus follows that
\begin{align}
	\alpha_{m-2}=-\binom{m}{m-2}+\binom{m}{m-1}\binom{m-1}{m-2}\,. \nonumber
\end{align}
Similar considerations lead to 
\begin{align}
	\alpha_{m-3} &= -\binom{m}{m-3}+\binom{m}{m-1}\binom{m-1}{m-3} + \binom{m}{m-2}\binom{m-2}{m-3} \nonumber\\ 
	&\quad+\binom{m}{m-1}\binom{m-1}{m-2}\binom{m-2}{m-3}\,.\nonumber
\end{align}
Each coefficient $\alpha_{\indA}$ includes one or several products of binomial coefficients, and each of these products has the form 
\begin{align}
	\binom{m}{m_1}\binom{m_1}{m_2}\dots\binom{m_{\indB-2}}{m_{\indB-1}}\binom{m_{\indB-1}}{\indA}\nonumber
\end{align}
for some $\indB\leq m-\indA$. Such a product of binomial coefficients corresponds to the number of ways that one can partition the edge set of $s$ into $\indB$ subsets with sizes 
\begin{align}
	m - m_1,\, m_1-m_2,\,\dots,\, m_{\indB-2}-m_{\indB-1},\, m_{\indB-1}-\indA,\, \indA\,.\nonumber
\end{align} 
The coefficient $\alpha_{\indA}$ includes one such term for each integer composition of $m$ that includes $\indA$. It follows that
\begin{align}
	\alpha_{\indA}=\sum_{{\bf q}\in\mathcal Q(m-\indA)} (-1)^{|{\bf q}|} \binom{m}{\indA}{\bf q}\tilde{!}\,,\label{eq:alphas}
\end{align}
where $\mathcal Q(m-\indA)$ is the set of integer compositions ${\bf q}=(q_1,q_2,\dots,q_{\indB})$ of $m-\indA$ and ${\bf q}\tilde{!}$ is the multinomial coefficient for the sequence $(q_1,q_2,\dots,q_{\indB})$ of $\indB$ integers. We use $|{\bf q}|:=k$ to denote the number of integers in an integer composition ${\bf q}$. Substituting the coefficients $\alpha_\indA$ into \eq{alphasum} using \eq{alphas} yields \eq{direct}.


\section{Derivation of the covariance matrix}\label{app:cov_deriv}
At time $t+dt$, the state vector of the mOUP with adjacency matrix $\bf A$, coupling parameter $\epsilon$, noise strength $\varsigma^2$, and reversion rate $\theta$ is
\begin{align}
	{\bf x}_{t+dt}={\bf K}{\bf x}_t+\varsigma\, dW_t\,,\label{eq:xtdt}
\end{align}
where ${\bf K}={\bf I}+\theta(\epsilon{\bf A}-{\bf I})dt$.

At steady state, the mOUP has the covariance matrix
\begin{align}
	\covmat=\langle {\bf x}_t\,{\bf x}_t^T\rangle&=\langle {\bf x}_{t+dt}\,{\bf x}_{t+dt}^T\rangle\,,\label{eq:varmatrix}
\end{align}
where $\langle\cdot\rangle$ denotes an ensemble average. We use \eq{xtdt} and substitute ${\bf x}_{t+dt}$ into \eq{varmatrix} to obtain
\begin{align}
	\covmat =\langle({\bf K}{\bf x}_t+\varsigma\, dW_t)({\bf K}{\bf x}_t+\varsigma\, dW_t)^T\rangle = \langle{\bf K}{\bf x}_t{\bf x}_t^T{\bf K}^T+\varsigma^2{\bf I}\,dt\rangle\,,\label{eq:kxxk}
\end{align}
where the second equality follows from the fact that $dW_t$ is a mean-0, unit-variance stochastic process that is independent of ${\bf x}_t$. 
Evaluating the ensemble average in \eq{kxxk} yields
\begin{align}
	\covmat&={\bf K}\covmat{\bf K}^T+\varsigma^2{\bf I}\, dt\nonumber\\
	&=[{\bf I}+\theta(\epsilon{\bf A}-{\bf I})\, dt]\covmat[{\bf I}+\theta(\epsilon{\bf A}-{\bf I})\, dt]^T + \varsigma^2{\bf I}\, dt\nonumber\\
	&=\covmat+\theta\left[(\epsilon{\bf A}-{\bf I})\covmat+\covmat
(\epsilon{\bf A}-{\bf I})^T+\frac{\varsigma^2}{\theta}{\bf I}\right]dt + O((dt)^2)\,.\nonumber
\end{align}
To first order in $dt$, we thus have
\begin{align}
	0=(\epsilon{\bf A}-{\bf I})\covmat+\covmat
(\epsilon{\bf A}-{\bf I})^T+\frac{\varsigma^2}{\theta}{\bf I}\,.\label{eq:l1}
\end{align}
Equation (\ref{eq:l1}) is a Lyapunov equation \cite{Chen2013, Fairman1998}. For the mOUP with signal decay, the solution of \eq{l1} is \cite{Fairman1998}
\begin{align}
	\covmat &= \frac{\varsigma^2}{\theta}\int_0^\infty e^{(\epsilon{\bf A}-{\bf I})t}e^{(\epsilon{\bf A}^T-{\bf I})t}\,dt=\frac{\varsigma^2}{\theta}\covmat_0\,,\label{eq:gramian}
\end{align}
where 
\begin{align}
	\covmat_0:=\int_0^\infty e^{(\epsilon{\bf A}-{\bf I})t}e^{(\epsilon{\bf A}^T-{\bf I})t}\,dt\nonumber
\end{align}
 is the covariance matrix of the mOUP when ${\varsigma=\theta=1}$. For $\varsigma=\theta=1$, Barnett et al.~\cite{Barnett2009,Barnett2011} derived the covariance matrix as a sum of products of $\bf A$ and ${\bf A}^T$ to yield the equation\footnote{In Barnett et al.~\cite{Barnett2009, Barnett2011}, the order of $\bf A$ and ${\bf A}^T$ is reversed, because they used row vectors instead of column vectors to describe the state ${\bf x}_t$.}:
\begin{align}
	\covmat_0 = \sum_{L=0}^\infty 2^{-(L+1)} \sum_{\ell=0}^L \binom{L}{\ell} (\epsilon {\bf A})^\ell (\epsilon {\bf A}^T)^{L-\ell}\,.
\label{eq:v0}
\end{align}
Therefore,
\begin{align}
	\covmat = \frac{\varsigma^2}{\theta}\sum_{L=0}^\infty 2^{-(L+1)} \sum_{\ell=0}^L \binom{L}{\ell} (\epsilon {\bf A})^\ell (\epsilon {\bf A}^T)^{L-\ell}\,.\label{eq:vsum2}
\end{align}


\section{Conditions for short-range signal decay}\label{app:short_range}
The sums in \eq{correl1} converge if the matrix $\covmat_0=\frac{\theta}{\varsigma^2}\covmat$ has eigenvalues $\nu_{i=1,\dots,n}\in(0,1)$ \cite[p.~38]{Hall2015}. The covariance matrix $\covmat_0$ is a symmetric, positive-semidefinite matrix. Therefore, a sufficient condition for short-range signal decay only needs to constrain the largest eigenvalue of $\covmat_0$.

First, we show that a sufficient condition for short-range signal decay is 
\begin{align}
	\|\epsilon {\bf A}\|_2<\frac{1}{2}\,.\label{eq:hscond}
\end{align}
Applying the Hilbert--Schmidt norm to both sides of \eq{v0} yields
\begin{align}
	\|\covmat_0\|_2 &= \norm{\sum_{L=0}^\infty 2^{-(L+1)} \sum_{\ell=0}^l \binom{L}{\ell} (\epsilon {\bf A}^T)^\ell (\epsilon {\bf A})^{L-\ell}}_2\nonumber\\
&\leq \sum_{L=0}^\infty 2^{-(L+1)} \sum_{\ell=0}^L \binom{L}{\ell} \|\epsilon {\bf A}\|_2^L\nonumber\\
&=\frac{1}{2}\sum_{L=0}^\infty \|\epsilon {\bf A}\|_2^L\,,\nonumber
\end{align}
where we used the identity $\|{\bf A}^T\|_2=\|{\bf A}\|_2$ and subadditivity and submultiplicativity of the Hilbert--Schmidt norm. 
When $\|\epsilon {\bf A}\|_2<1/2$, it follows that $\|\covmat_0\|_2<1$, so $\nu_{i=1,\dots,n}\in(0,1)$ for the positive-semidefinite matrix $\covmat_0$. It follows that the sums in \eq{correl1} converge.

For many applications in network analysis, the spectral radius $\rho(\cdot)$ (which is equal to the largest absolute value of the eigenvalues of a matrix) is a commonly used matrix norm \cite{Jamakovic2006, Stevanovic2014}. We now show that one can relax the condition in \eq{hscond} for short-range signal decay to
\begin{align}
	\rho(\epsilon{\bf A})<\frac{1}{2}
\end{align}
if ${\bf A}$ is the adjacency matrix of a strongly connected graph with non-negative edge weights.

The adjacency matrix of a strongly connected graph is irreducible \cite{Boyle2015}. For an irreducible matrix with non-negative entries, the Perron--Frobenius theorem guarantees the existence of a simple, real, positive eigenvalue $\lambda_\textrm{max}=\rho({\bf A})$ \cite{Boyle2015}. The transpose of ${\bf A}$ is also an adjacency matrix of a strongly connected graph with non-negative edge weights, so ${\bf A}^T$ also has a simple, positive, real leading eigenvalue. Ortega \cite[p.~24]{Ortega1990} proved the existence of a submultiplicative matrix norm $\|{\bf M}\|$, such that $\rho({\bf M})=\|{\bf M}\|$ for all complex square matrices $\bf M$ with simple max-modulus eigenvalues\footnote{A max-modulus eigenvalue $\mu$ of a matrix ${\bf M}$ is an eigenvalue that satisfies $\mu=\rho({\bf M})$.}. The matrices ${\bf A}$ and ${\bf A}^T$ are matrices with a single simple max-modulus eigenvalue. We thus write
\begin{align}
	\rho(\covmat_0) &\leq \|\covmat_0\| \nonumber\\&\leq \norm{\sum_{L=0}^\infty 2^{-(L+1)} \sum_{\ell=0}^l \binom{L}{\ell} (\epsilon {\bf A}^T)^\ell (\epsilon {\bf A})^{L-\ell}}\nonumber
\end{align}
and use the subadditivity and submulitplicativity of $\|\cdot\|$ to obtain
\begin{align}
	\rho(\covmat) \leq \frac{1}{2} \sum_{L=0}^\infty \|\epsilon {\bf A}\|^L=\frac{1}{2} \sum_{L=0}^\infty \left(\rho(\epsilon {\bf A})\right)^L\,.\label{eq:rhoA}
\end{align}
When $\rho(\epsilon{\bf A})<1/2$, it follows from \eq{rhoA} that $\rho(\covmat_0)<1$. It then follows that $\nu_{i=1,\dots,n}\in(0,1)$, so the sums in \eq{correl1} converge.


\section*{Acknowledgements}

We thank Alex Arenas, Lionel Barnett, Heather Zinn Brooks, Bing Brunton, Michelle Feng, Kameron Decker Harris, Renaud Lambiotte, Neave O'Cleary, Gesine Reinert, and Jonny Wray for helpful discussions and comments. We also thank our reviewers for their helpful comments and suggestions.



\begin{thebibliography}{100}

\bibitem{SInote}
The Jupyter notebook is available as a web application under
  \url{https://gitlab.com/aliceschwarze/motifs-for-processes}.

\bibitem{Aalen2004}
{\sc O.~O. Aalen and H.~K. Gjessing}, {\em Survival models based on the
  {Ornstein}--{Uhlenbeck} process}, {Lifetime Data Analysis}, 10 (2004),
  pp.~407--423.

\bibitem{Alon2011}
{\sc N.~Alon, C.~Avin, M.~Kouck{\`y}, G.~Kozma, Z.~Lotker, and M.~R. Tuttle},
  {\em {Many random walks are faster than one}}, {Combinatorics, Probability
  and Computing}, 20 (2011), pp.~481--502.

\bibitem{Alon2007}
{\sc U.~Alon}, {\em {Network motifs: Theory and experimental approaches}},
  {Nature Reviews Genetics}, 8 (2007), pp.~450--461.

\bibitem{Antoneli2018}
{\sc F.~Antoneli, M.~Golubitsky, and I.~Stewart}, {\em {Homeostasis in a feed
  forward loop gene regulatory motif}}, {Journal of Theoretical Biology}, 445
  (2018), pp.~103--109.

\bibitem{Artzy-Randrup2004}
{\sc Y.~Artzy-Randrup, S.~J. Fleishman, N.~Ben-Tal, and L.~Stone}, {\em
  {Comment on ``Network motifs: simple building blocks of complex networks''
  and ``Superfamilies of evolved and designed networks''}}, {Science}, 305
  (2004), pp.~1107--1107.

\bibitem{Barnett2009}
{\sc L.~Barnett, C.~L. Buckley, and S.~Bullock}, {\em {Neural complexity and
  structural connectivity}}, {Physical Review E}, 79 (2009), p.~051914.

\bibitem{Barnett2011}
{\sc L.~Barnett, C.~L. Buckley, and S.~Bullock}, {\em {Neural complexity: A
  graph theoretic interpretation}}, {Physical Review E}, 83 (2011), p.~041906.

\bibitem{Barzel2009}
{\sc B.~Barzel and O.~Biham}, {\em Quantifying the connectivity of a network:
  {T}he network correlation function method}, Physical Review E, 80 (2009),
  p.~046104.

\bibitem{Battiston2017}
{\sc F.~Battiston, V.~Nicosia, M.~Chavez, and V.~Latora}, {\em {Multilayer
  motif analysis of brain networks}}, {Chaos: An Interdisciplinary Journal of
  Nonlinear Science}, 27 (2017), p.~047404.

\bibitem{Bianconi2004}
{\sc G.~Bianconi}, {\em {Number of cycles in off-equilibrium scale-free
  networks and in the internet at the autonomous system level}}, {The European
  Physical Journal B}, 38 (2004), pp.~223--230.

\bibitem{Bianconi2017}
{\sc G.~Bianconi}, {\em {Epidemic spreading and bond percolation on multilayer
  networks}}, {Journal of Statistical Mechanics: Theory and Experiment}, 2017
  (2017), p.~034001.

\bibitem{Bollobas2013}
{\sc B.~Bollob{\'a}s}, {\em Modern {G}raph {T}heory}, vol.~184, Springer,
  {Berlin, Germany}, 2013.

\bibitem{Boergers2017}
{\sc C.~B{\"o}rgers}, {\em {An Introduction to Modeling Neuronal Dynamics}},
  Springer, {Cham, Switzerland}, 2017.

\bibitem{Boyle2015}
{\sc M.~Boyle}, {\em Notes on the {Perron--Frobenius} theory of nonnegative
  matrices}, 2015,
  \url{https://www.math.umd.edu/~mboyle/courses/405sp10/specmay2011.pdf}.

\bibitem{Carter2004}
{\sc S.~L. Carter, C.~M. Brechb{\"u}hler, M.~Griffin, and A.~T. Bond}, {\em
  {Gene co-expression network topology provides a framework for molecular
  characterization of cellular state}}, {Bioinformatics}, 20 (2004),
  pp.~2242--2250.

\bibitem{Chandra2020}
{\sc S.~Chandra, E.~Ott, and M.~Girvan}, {\em {Critical network cascades with
  re-excitable nodes: Why treelike approximations usually work, when they break
  down, and how to correct them}}, {Physical Review E}, 101 (2020), p.~062304.

\bibitem{Chen2013}
{\sc C.-T. Chen}, {\em {Linear System Theory and Design}}, Oxford University
  Press, Inc., Oxford, United Kingdom, 4th~ed., 2013.

\bibitem{Ciriello2008}
{\sc G.~Ciriello and C.~Guerra}, {\em {A review on models and algorithms for
  motif discovery in protein--protein interaction networks}}, {Briefings in
  Functional Genomics and Proteomics}, 7 (2008), pp.~147--156.

\bibitem{Conant2003}
{\sc G.~C. Conant and A.~Wagner}, {\em {Convergent evolution of gene
  circuits}}, {Nature Genetics}, 34 (2003), pp.~264--266.

\bibitem{Dechery2018}
{\sc J.~B. Dechery and J.~N. MacLean}, {\em {Functional triplet motifs underlie
  accurate predictions of single-trial responses in populations of tuned and
  untuned V1 neurons}}, {PLoS Computational Biology}, 14 (2018), p.~e1006153.

\bibitem{Demirel2014}
{\sc G.~Demirel, F.~Vazquez, G.~B{\"o}hme, and T.~Gross}, {\em {Moment-closure
  approximations for discrete adaptive networks}}, {Physica D: Nonlinear
  Phenomena}, 267 (2014), pp.~68--80.

\bibitem{Eger2013}
{\sc S.~Eger}, {\em Restricted weighted integer compositions and extended
  binomial coefficients}, {Journal of Integer Sequences}, 16 (2013), p.~3.

\bibitem{Estrada2005}
{\sc E.~Estrada and J.~A. Rodriguez-Velazquez}, {\em {Subgraph centrality in
  complex networks}}, {Physical Review E}, 71 (2005), p.~056103.

\bibitem{Fairman1998}
{\sc F.~W. Fairman}, {\em {Linear Control Theory: The State Space Approach}},
  John Wiley \& Sons, New York City, NY, USA, 1998.

\bibitem{Fox2010}
{\sc M.~D. Fox and M.~Greicius}, {\em {Clinical applications of resting state
  functional connectivity}}, {Frontiers in Systems Neuroscience}, 4 (2010).

\bibitem{Garcia2012}
{\sc G.~C. Garcia, A.~Lesne, M.-T. H{\"u}tt, and C.~C. Hilgetag}, {\em
  {Building blocks of self-sustained activity in a simple deterministic model
  of excitable neural networks}}, {Frontiers in Computational Neuroscience}, 6
  (2012), p.~50.

\bibitem{Gleiss2001}
{\sc P.~M. Gleiss, P.~F. Stadler, A.~Wagner, and D.~A. Fell}, {\em {Relevant
  cycles in chemical reaction networks}}, {Advances in Complex Systems}, 4
  (2001), pp.~207--226.

\bibitem{Godsil2011}
{\sc C.~Godsil and K.~Guo}, {\em {Quantum walks on regular graphs and
  eigenvalues}}, {The Electronic Journal of Combinatorics}, 18 (2011), p.~P165.

\bibitem{Golubitsky2009}
{\sc M.~Golubitsky, L.~Shiau, C.~Postlethwaite, and Y.~Zhang}, {\em The
  feed-forward chain as a filter-amplifier motif}, in Coherent Behavior in
  Neuronal Networks, Springer-Verlag, Heidelberg, Germany, 2009, pp.~95--120.

\bibitem{Grytskyy2013}
{\sc D.~Grytskyy, T.~Tetzlaff, M.~Diesmann, and M.~Helias}, {\em {A unified
  view on weakly correlated recurrent networks}}, {Frontiers in Computational
  Neuroscience}, 7 (2013), p.~131.

\bibitem{Hall2015}
{\sc B.~Hall}, {\em {Lie Groups, Lie Algebras, and Representations: An
  Elementary Introduction}}, vol.~222, Springer-Verlag, {Cham, Switzerland},
  2015.

\bibitem{Hawkes1971}
{\sc A.~G. Hawkes}, {\em Point spectra of some mutually exciting point
  processes}, {Journal of the Royal Statistical Society: Series B
  (Methodological)}, 33 (1971), pp.~438--443.

\bibitem{Haydon2003}
{\sc D.~T. Haydon, M.~Chase-Topping, D.~J. Shaw, L.~Matthews, J.~K. Friar,
  J.~Wilesmith, and M.~E.~J. Woolhouse}, {\em The construction and analysis of
  epidemic trees with reference to the 2001 uk foot--and--mouth outbreak},
  {Proceedings of the Royal Society of London. Series B: Biological Sciences},
  270 (2003), pp.~121--127.

\bibitem{Holme2019}
{\sc P.~Holme and J.~Saram{\"a}ki}, {\em {Temporal Network Theory}}, Springer,
  {Cham, Switzerland}, 2019.

\bibitem{Hong-Lin2014}
{\sc X.~{Hong-Lin}, Y.~{Han-Bing}, G.~{Cui-Fang}, and Z.~Ping}, {\em Social
  network analysis based on network motifs}, {Journal of Applied Mathematics},
  2014 (2014).

\bibitem{House2009}
{\sc T.~House, G.~Davies, L.~Danon, and M.~J. Keeling}, {\em {A motif-based
  approach to network epidemics}}, {Bulletin of Mathematical Biology}, 71
  (2009), pp.~1693--1706.

\bibitem{Hu2014}
{\sc Y.~Hu, J.~Trousdale, K.~Josi{\'c}, and E.~Shea-Brown}, {\em {Local paths
  to global coherence: Cutting networks down to size}}, {Physical Review E}, 89
  (2014), p.~032802.

\bibitem{Hulovatyy2015}
{\sc Y.~Hulovatyy, H.~Chen, and T.~Milenkovi{\'c}}, {\em Exploring the
  structure and function of temporal networks with dynamic graphlets},
  {Bioinformatics}, 31 (2015), pp.~i171--i180.

\bibitem{Ingram2006}
{\sc P.~J. Ingram, M.~P. Stumpf, and J.~Stark}, {\em {Network motifs: Structure
  does not determine function}}, {BMC Genomics}, 7 (2006), p.~108.

\bibitem{Iturria-Medina2008}
{\sc Y.~Iturria-Medina, R.~C. Sotero, E.~J. Canales-Rodr{\'\i}guez,
  Y.~Alem{\'a}n-G{\'o}mez, and L.~Melie-Garc{\'\i}a}, {\em {Studying the human
  brain anatomical network via diffusion-weighted MRI and Graph Theory}},
  {Neuroimage}, 40 (2008), pp.~1064--1076.

\bibitem{Jamakovic2006}
{\sc A.~Jamakovic, R.~Kooij, P.~Van~Mieghem, and E.~R. van Dam}, {\em
  {Robustness of networks against viruses: The role of the spectral radius}},
  in {IEEE 2006 Symposium on Communications and Vehicular Technology}, 2006,
  pp.~35--38.

\bibitem{Jovanovic2016}
{\sc S.~Jovanovi{\'c} and S.~Rotter}, {\em {Interplay between graph topology
  and correlations of third order in spiking neuronal networks}}, {PLoS
  Computational Biology}, 12 (2016).

\bibitem{Juszczyszyn2012}
{\sc K.~Juszczyszyn, K.~Musia{\l}, P.~Kazienko, and B.~Gabrys}, {\em Temporal
  changes in local topology of an email-based social network}, {Computing and
  Informatics}, 28 (2012), pp.~763--779.

\bibitem{Juul2020}
{\sc J.~S. Juul and S.~H. Strogatz}, {\em Descendant distributions for the
  impact of mutant contagion on networks}, {Physical Review Research}, 2
  (2020), p.~033005.

\bibitem{Kampa2006}
{\sc B.~M. Kampa, J.~J. Letzkus, and G.~J. Stuart}, {\em Cortical feed-forward
  networks for binding different streams of sensory information}, {Nature
  Neuroscience}, 9 (2006), pp.~1472--1473.

\bibitem{Kiss2017}
{\sc I.~Z. Kiss, J.~C. Miller, and P.~L. Simon}, {\em {Mathematics of Epidemics
  on Networks}}, Springer, {Cham, Switzerland}, 2017.

\bibitem{Larremore2012}
{\sc D.~B. Larremore, M.~Y. Carpenter, E.~Ott, and J.~G. Restrepo}, {\em
  {Statistical properties of avalanches in networks}}, {Physical Review E}, 85
  (2012), p.~066131.

\bibitem{Lehmann2019}
{\sc S.~Lehmann}, {\em Fundamental structures in dynamic communication
  networks}, {arXiv:1907.09966},  (2019).

\bibitem{Lehman2018}
{\sc S.~Lehmann and Y.-Y. Ahn}, {\em {Complex Spreading Phenomena in Social
  Systems}}, Springer, {Cham, Switzerland}, 2018.

\bibitem{Liang2011}
{\sc Z.~Liang, K.~C. Yuen, and J.~Guo}, {\em Optimal proportional reinsurance
  and investment in a stock market with {Ornstein--Uhlenbeck} process},
  {Insurance: Mathematics and Economics}, 49 (2011), pp.~207--215.

\bibitem{Liu2020}
{\sc P.~Liu, V.~Guarrasi, and A.~E. Sar{\i}y{\"u}ce}, {\em {Temporal network
  motifs: Models, limitations, evaluation}}, {arXiv:2005.11817},  (2020).

\bibitem{Lizier2012}
{\sc J.~T. Lizier, F.~M. Atay, and J.~Jost}, {\em {Information storage, loop
  motifs, and clustered structure in complex networks}}, {Physical Review E},
  86 (2012), p.~026110.

\bibitem{Maayan2008}
{\sc A.~Ma'ayan, G.~A. Cecchi, J.~Wagner, A.~R. Rao, R.~Iyengar, and
  G.~Stolovitzky}, {\em {Ordered cyclic motifs contribute to dynamic stability
  in biological and engineered networks}}, {Proceedings of the National Academy
  of Sciences of the United States of America}, 105 (2008), pp.~19235--19240.

\bibitem{Manjunath2013}
{\sc G.~Manjunath and H.~Jaeger}, {\em {Echo state property linked to an input:
  Exploring a fundamental characteristic of recurrent neural networks}},
  {Neural Computation}, 25 (2013), pp.~671--696.

\bibitem{Manrubia2003}
{\sc S.~Manrubia and J.~Poyatos}, {\em {Motif selection in a model of evolving
  replicators: The role of surfaces and limited transport in network
  topology}}, {Europhysics Letters}, 64 (2003), p.~557.

\bibitem{Masoudi-Nejad2012}
{\sc A.~Masoudi-Nejad, F.~Schreiber, and Z.~R.~M. Kashani}, {\em {Building
  blocks of biological networks: A review on major network motif discovery
  algorithms}}, {Iet Systems Biology}, 6 (2012), pp.~164--174.

\bibitem{Melnik2011}
{\sc S.~Melnik, A.~Hackett, M.~A. Porter, P.~J. Mucha, and J.~P. Gleeson}, {\em
  The unreasonable effectiveness of tree-based theory for networks with
  clustering}, {Physical Review E}, 83 (2011), p.~036112.

\bibitem{Milo2002}
{\sc R.~Milo, S.~Shen-Orr, S.~Itzkovitz, N.~Kashtan, D.~Chklovskii, and
  U.~Alon}, {\em {Network motifs: Simple building blocks of complex networks}},
  {Science}, 298 (2002), pp.~824--827.

\bibitem{Newman2002}
{\sc M.~E.~J. Newman}, {\em {Spread of epidemic disease on networks}},
  {Physical Review E}, 66 (2002), p.~016128.

\bibitem{Newman2018}
{\sc M.~E.~J. Newman}, {\em Networks}, {Oxford University Press}, Oxford,
  United Kingdom, 2018.

\bibitem{Nguyen2018}
{\sc G.~H. Nguyen, J.~B. Lee, R.~A. Rossi, N.~K. Ahmed, E.~Koh, and S.~Kim},
  {\em {Continuous-time dynamic network embeddings}}, in {Companion Proceedings
  of the The Web Conference 2018}, 2018, pp.~969--976.

\bibitem{Novelli2020}
{\sc L.~Novelli, F.~M. Atay, J.~Jost, and J.~T. Lizier}, {\em Deriving pairwise
  transfer entropy from network structure and motifs}, {Proceedings of the
  Royal Society A}, 476 (2020), p.~20190779.

\bibitem{Ogbogbo2018}
{\sc C.~P. Ogbogbo}, {\em Modeling crude oil spot price as an
  {Ornstein--Uhlenbeck} process}, {International Journal of Mathematical
  Analysis and Optimization: Theory and Applications}, 2018 (2018),
  pp.~261--275.

\bibitem{Oh2018}
{\sc S.-W. Oh and M.~A. Porter}, {\em Complex contagions with timers}, {Chaos:
  An Interdisciplinary Journal of Nonlinear Science}, 28 (2018), p.~033101.

\bibitem{Ohnishi2010}
{\sc T.~Ohnishi, H.~Takayasu, and M.~Takayasu}, {\em {Network motifs in an
  inter-firm network}}, {Journal of Economic Interaction and Coordination}, 5
  (2010), pp.~171--180.

\bibitem{Onnela2004}
{\sc J.-P. Onnela, K.~Kaski, and J.~Kert{\'e}sz}, {\em {Clustering and
  information in correlation based financial networks}}, {The European Physical
  Journal B}, 38 (2004), pp.~353--362.

\bibitem{Ortega1990}
{\sc J.~M. Ortega}, {\em {Numerical Analysis: A Second Course}}, SIAM,
  Philadelphia, PA, USA, 1990.

\bibitem{Paranjape2017}
{\sc A.~Paranjape, A.~R. Benson, and J.~Leskovec}, {\em Motifs in temporal
  networks}, in {Proceedings of the Tenth ACM International Conference on Web
  Search and Data Mining}, 2017, pp.~601--610.

\bibitem{Pastor-Satorras2015}
{\sc R.~Pastor-Satorras, C.~Castellano, P.~Van~Mieghem, and A.~Vespignani},
  {\em Epidemic processes in complex networks}, {Reviews of Modern Physics}, 87
  (2015), p.~925.

\bibitem{Pernice2011}
{\sc V.~Pernice, B.~Staude, S.~Cardanobile, and S.~Rotter}, {\em How structure
  determines correlations in neuronal networks}, {PLoS Computational Biology},
  7 (2011), p.~e1002059.

\bibitem{Porter2016}
{\sc M.~A. Porter and J.~P. Gleeson}, {\em {Dynamical Systems on Networks}},
  {Frontiers in Applied Dynamical Systems: Reviews and Tutorials}, 4 (2016).

\bibitem{Radicchi2016}
{\sc F.~Radicchi and C.~Castellano}, {\em {Leveraging percolation theory to
  single out influential spreaders in networks}}, {Physical Review E}, 93
  (2016), p.~062314.

\bibitem{Reichenbach1956}
{\sc H.~Reichenbach}, {\em The Direction of Time}, University of California
  Press, Berkeley and Los Angeles, CA, USA, 1956.

\bibitem{Rice2006}
{\sc J.~Rice}, {\em {Mathematical Statistics and Data Analysis}}, Thomson
  Higher Education, Belmont, CA, USA, 3rd~ed., 2006.

\bibitem{Rip2010}
{\sc J.~M. Rip, K.~S. McCann, D.~H. Lynn, and S.~Fawcett}, {\em {An
  experimental test of a fundamental food web motif}}, {Proceedings of the
  Royal Society of London B: Biological Sciences}, 277 (2010), pp.~1743--1749.

\bibitem{Ristl2014}
{\sc K.~Ristl, S.~J. Plitzko, and B.~Drossel}, {\em Complex response of a
  food-web module to symmetric and asymmetric migration between several
  patches}, {Journal of Theoretical Biology}, 354 (2014), pp.~54--59.

\bibitem{Robin2007}
{\sc S.~Robin, S.~Schbath, and V.~Vandewalle}, {\em Statistical tests to
  compare motif count exceptionalities}, {BMC Bioinformatics}, 8 (2007), p.~84.

\bibitem{Rohlfs2013}
{\sc R.~V. Rohlfs, P.~Harrigan, and R.~Nielsen}, {\em Modeling gene expression
  evolution with an extended {Ornstein--Uhlenbeck} process accounting for
  within-species variation}, {Molecular Biology and Evolution}, 31 (2013),
  pp.~201--211.

\bibitem{Roldan2020}
{\sc J.~M. Roldan, V.~K. George, G.~A. Silva, et~al.}, {\em Construction of
  edge-ordered multidirected graphlets for comparing dynamics of spatial
  temporal neural networks}, {arXiv:2006.15971},  (2020).

\bibitem{Sanderson2016}
{\sc T.~Sanderson, G.~Hertzler, T.~Capon, and P.~Hayman}, {\em A real options
  analysis of australian wheat production under climate change}, {Australian
  Journal of Agricultural and Resource Economics}, 60 (2016), pp.~79--96.

\bibitem{Schlauch2015}
{\sc W.~E. Schlauch and K.~A. Zweig}, {\em {Influence of the null-model on
  motif detection}}, in {2015 IEEE/ACM International Conference on Advances in
  Social Networks Analysis and Mining (ASONAM)}, IEEE, 2015, pp.~514--519.

\bibitem{Shen2012}
{\sc K.~Shen, G.~Bezgin, R.~M. Hutchison, J.~S. Gati, R.~S. Menon, S.~Everling,
  and A.~R. McIntosh}, {\em {Information processing architecture of
  functionally defined clusters in the macaque cortex}}, {Journal of
  Neuroscience}, 32 (2012), pp.~17465--17476.

\bibitem{Shen-Orr2002}
{\sc S.~S. Shen-Orr, R.~Milo, S.~Mangan, and U.~Alon}, {\em Network motifs in
  the transcriptional regulation network of \emph{Escherichia coli}}, {Nature
  Genetics}, 31 (2002), pp.~64--68.

\bibitem{Shilnikov2008}
{\sc A.~Shilnikov, R.~Gordon, and I.~Belykh}, {\em {Polyrhythmic
  synchronization in bursting networking motifs}}, {Chaos: An Interdisciplinary
  Journal of Nonlinear Science}, 18 (2008), p.~037120.

\bibitem{Shinozaki2015}
{\sc T.~Shinozaki and S.~Watanabe}, {\em Structure discovery of deep neural
  network based on evolutionary algorithms}, in {2015 IEEE International
  Conference on Acoustics, Speech and Signal Processing (ICASSP)}, 2015,
  pp.~4979--4983.

\bibitem{Spielman2012}
{\sc D.~Spielman}, {\em {Spectral graph theory}}, in {C}ombinatorial
  {S}cientific {C}omputing, CRC Press, Boca Raton, Florida, USA, 2012.

\bibitem{Sporns2004}
{\sc O.~Sporns and R.~K{\"o}tter}, {\em {Motifs in brain networks}}, {PLoS
  Biology}, 2 (2004), p.~e369.

\bibitem{Sporns2000a}
{\sc O.~Sporns, G.~Tononi, and G.~M. Edelman}, {\em {Connectivity and
  complexity: The relationship between neuroanatomy and brain dynamics}},
  {Neural Networks}, 13 (2000), pp.~909--922.

\bibitem{Sporns2000}
{\sc O.~Sporns, G.~Tononi, and G.~M. Edelman}, {\em {Theoretical neuroanatomy:
  Relating anatomical and functional connectivity in graphs and cortical
  connection matrices}}, {Cerebral Cortex}, 10 (2000), pp.~127--141.

\bibitem{Stevanovic2014}
{\sc D.~Stevanovic}, {\em {Spectral Radius of Graphs}}, {Academic Press},
  London, UK, 2014.

\bibitem{Stone2019}
{\sc L.~Stone, D.~Simberloff, and Y.~Artzy-Randrup}, {\em {Network motifs and
  their origins}}, {PLoS Computational Biology}, 15 (2019), p.~e1006749.

\bibitem{Takes2018}
{\sc F.~W. Takes, W.~A. Kosters, B.~Witte, and E.~M. Heemskerk}, {\em Multiplex
  network motifs as building blocks of corporate networks}, {Applied Network
  Science}, 3 (2018), p.~39.

\bibitem{Tino2015}
{\sc P.~Tino, L.~Benuskova, and A.~Sperduti}, {\em Artificial neural network
  models}, in {Springer Handbook of Computational Intelligence}, Springer,
  Cham, Switzerland, 2015, pp.~455--471.

\bibitem{Tononi1994}
{\sc G.~Tononi, O.~Sporns, and G.~M. Edelman}, {\em {A measure for brain
  complexity: Relating functional segregation and integration in the nervous
  system}}, {Proceedings of the National Academy of Sciences of the United
  States of America}, 91 (1994), pp.~5033--5037.

\bibitem{Trousdale2012}
{\sc J.~Trousdale, Y.~Hu, E.~Shea-Brown, and K.~Josi{\'c}}, {\em {Impact of
  network structure and cellular response on spike time correlations}}, {PLoS
  Computational Biology}, 8 (2012), p.~e1002408.

\bibitem{Trudeau2013}
{\sc R.~J. Trudeau}, {\em {Introduction to Graph Theory}}, Dover Publications,
  New York, NY, USA, 2013.

\bibitem{Tsai2011}
{\sc M.-T. Tsai, J.-D. Saphores, and A.~Regan}, {\em Valuation of freight
  transportation contracts under uncertainty}, {Transportation Research Part E:
  Logistics and Transportation Review}, 47 (2011), pp.~920--932.

\bibitem{Tu2018}
{\sc K.~Tu, J.~Li, D.~Towsley, D.~Braines, and L.~D. Turner}, {\em Network
  classification in temporal networks using motifs}, {arXiv:1807.03733},
  (2018).

\bibitem{Uhlenbeck1930}
{\sc G.~E. Uhlenbeck and L.~S. Ornstein}, {\em On the theory of the {B}rownian
  motion}, {Physical Review}, 36 (1930), p.~823.

\bibitem{Vanrullen2003}
{\sc R.~{VanRullen} and C.~Koch}, {\em Visual selective behavior can be
  triggered by a feed-forward process}, {Journal of Cognitive Neuroscience}, 15
  (2003), pp.~209--217.

\bibitem{Vazquez2004}
{\sc A.~Vazquez, R.~Dobrin, D.~Sergi, J.-P. Eckmann, Z.~N. Oltvai, and A.-L.
  Barab{\'a}si}, {\em {The topological relationship between the large-scale
  attributes and local interaction patterns of complex networks}}, {Proceedings
  of the National Academy of Sciences of the United States of America}, 101
  (2004), pp.~17940--17945.

\bibitem{Wu2012}
{\sc G.~Wu, M.~Harrigan, and P.~Cunningham}, {\em Classifying wikipedia
  articles using network motif counts and ratios}, in {Proceedings of the
  Eighth Annual International Symposium on Wikis and Open Collaboration},
  no.~12, 2012, pp.~1--10.

\bibitem{Yaveroglu2014}
{\sc {\"O}.~N. Yavero{\u{g}}lu, N.~Malod-Dognin, D.~Davis, Z.~Levnajic,
  V.~Janjic, R.~Karapandza, A.~Stojmirovic, and N.~Pr{\v{z}}ulj}, {\em
  Revealing the hidden language of complex networks}, {Scientific Reports}, 4
  (2014), p.~4547.

\bibitem{Zhigulin2004}
{\sc V.~P. Zhigulin}, {\em {Dynamical motifs: Building blocks of complex
  dynamics in sparsely connected random networks}}, {Physical Review Letters},
  92 (2004), p.~238701.

\end{thebibliography}



\end{document}